# One-batch Preempt Deterioration-effect Multi-state Multi-rework Network Reliability Problem and Algorithms

Wei-Chang Yeh
Integration & Collaboration Laboratory
Department of Industrial Engineering and Management Engineering
National Tsing Hua University, Hsinchu, Taiwan
yeh@ieee.org

Abstract – A rework network is a distinct multi-state network that exists in many real-life industrial manufacturing systems for fixing defective products using rework processes to improve the utility and productivity of the systems. To provide a more general model of rework networks, a novel one-batch preempt multi-state multi-rework network is proposed to achieve real-life applications without numerous impractical and unreasonable limitations. Accordingly, a new algorithm based on two different multi-state types of binary addition tree algorithms (BATs) is proposed to calculate the reliability of the proposed rework. Furthermore, the proposed BAT-based algorithm is applied to four rework problems to validate its effectiveness and performance for the rework reliability problem.

## 1. INTRODUCTION

Network reliability is a distinct success probability that plays an important role in the design, performance, and management of real-life systems and modern networks, e.g., Internet of things [1, 2], 4G/5G [3, 4], cloud computing [5, 6], smart grid [5, 7, 8], traffic networks [9, 10], and water-pipe networks [11, 12].

A rework network, $G(V, E, \mathbf{D})$, is a unique network including the deterioration effect and at least one rework process, where the work in products (WIP) is processed in a sequence of nodes, frequently resulting in defective products. Each node in the node set, $V = \{1, 2, …, n\}$, can be either a machine, process station, workstation, tool, unit, or worker for processing (including any type of manufacture) the WIPs that are input to the rework network. Each arc in the set, $E = \{e_{i,j} = a_k \mid i, j \in$ $V$}, connects a pair of consecutive nodes with the deterioration effect, and $\mathbf{D} = \{ (\mathbf{D}_i, \mathbf{D}(i, l)) \mid$ for all $i \in V\}$ is the state distribution of which $\mathbf{D}_i = \{0, 1, \ldots, \mathbf{D}_{max}(i)\}$, $\mathbf{D}_{max}(i)$ is maximal state level of node $i$, and $\mathbf{D}(i, l)$ is the occurrent probability of the related state level $l$ of node $i$.

For example, a single-rework network, $G(V, E, \mathbf{D})$, with only one rework process is depicted in Fig. 1, where $V = \{1, 2\}$, $E = \{a_1 = a_{1,1}, a_2 = a_{3,1}, \ldots, a_5 = a_{2,3}\}$, the rework process starts from nodes 2 to 1 to 2, and the state distribution, $\mathbf{D}$ is as summarized in Table 1 of which $\mathbf{D}(i, l)$ is the probability of node $i$ with state $l$ and $\mathbf{D}(a_j, l)$ is the probability of arc $a_j$ with state $l$, respectively.

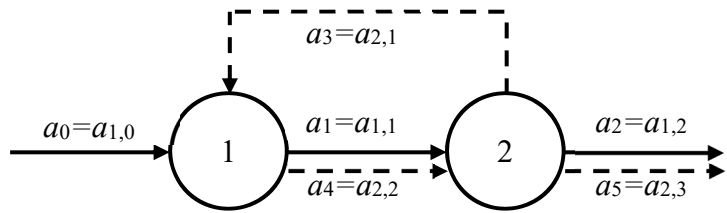

**Figure 1.** Example single-rework network.

**Table 1.** State distribution **D** of Fig. 1.

| State $l$ | $\mathbf{D}(1, l)$ | $\mathbf{D}(2, l)$ | $\mathbf{D}(a_0)$ | $\mathbf{D}(a_1)$ | $\mathbf{D}(a_2)$ | $\mathbf{D}(a_3)$ | $\mathbf{D}(a_4)$ | $\mathbf{D}(a_5)$ |
|---|---|---|---|---|---|---|---|---|
| 0 | 0.002 | 0.003 | 0.99 | 0.9 | 0.8 | 0.95 | 0.7 | 1.0 |
| 1 | 0.003 | 0.005 | | | | | | |
| 2 | 0.005 | 0.010 | | | | | | |
| 3 | 0.010 | 0.012 | | | | | | |
| 4 | 0.030 | 0.070 | | | | | | |
| 5 | 0.050 | 0.900 | | | | | | |
| 6 | 0.100 | | | | | | | |
| 7 | 0.800 | | | | | | | |

The deterioration effect is a distinct effect that decreases the WIPs in a production line from successive nodes owing to the discarded/defective products, and it causes the flow conservation law to be no longer held [13-17]. Hence, the number of WIPs is deceased after via arcs due to the deterioration effect. For example, three WIPs are input to node 1 and reduced to two WIPs immediately after leaving node 1 if one WIP is discarded.

Each WIP has three states: perfect, rework, and discarded. A perfect WIP is a WIP that has not been detected as a defect, i.e., it is perfect currently but could be a defect subsequently. A rework WIP is a WIP found to be defected after processing and not discarded in the rework process. After a perfect

WIP becomes defected, its state is changed to a rework WIP if it needs to be reworked or to a discarded WIP provided no rework is required.

In some practical applications, defect WIPs are extremely expensive to be discarded [14-19, 23, 27, 28]. After a defect WIP is detected, it is discarded if it is not detected in the rework node, e.g., node 2 in Fig. 1. Alternatively, it is sent to the rework process in $G(V, E, \mathbf{D})$ to be reworked to "as normal" condition. Note that a defect WIP can be discarded after the rework process fails.

A WIP cannot be changed to a perfect WIP once it becomes a rework WIP. A rework WIP is transformed to a discarded WIP if it has defected again (i.e., each WIP has at most one opportunity to be reworked) or sent to be processed in the rework network after it is fixed. The relationships among perfect, rework, and discarded WIPs are depicted in Fig. 2. Note that only the rework node has the mechanism to implement the rework process.

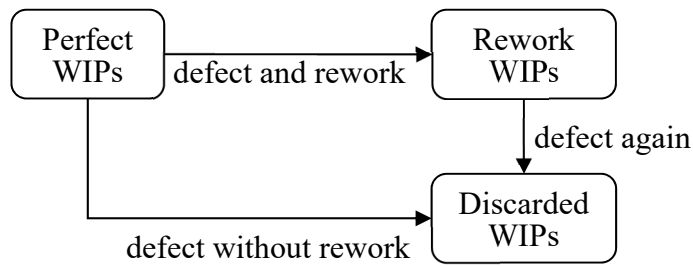

**Figure 2.** The relationship among perfect WIPs, rework WIPs, and discarded WIPs.

Without the loss of generality, each production line can have any WIP, such as oil, water, tiles, or cars [11, 12, 19-22], and it can be categorized as a perfect production line (e.g., the solid and dotted lines in Fig. 1) or many rework production lines (e.g., the dashed line in Fig. 1). The former has no rework WIPs, and the latter contain no perfect WIPs.

In this study, three common scenarios, one-batch, preempt, and the deterioration effect are adapted for real-life applications. The one-batch concept is that all the perfect WIPs need to wait for the rework WIPs to be reworked and then proceed to the next node together. For example, in Fig. 1, the perfect WIPs in node 2 must wait for the rework WIPs from node 2 to node 1 and return to node 2 to complete their rework process to leave node 2 together.

The preempt concept is that each WIP needs and uses a resource to be processed when it reaches a node irrespective of being discarded immediately after that node or sent to or returned from the rework process. Hence, a rework WIP requires extra capacity when it returns from the rework process. For example, in Fig. 1, assume a rework WIP consumes $a$ and $b$ units of capacity in node 2 before its rework process and on returning to node 2 after its rework process, respectively. Thus, for such a WIP in node 2, the total consumption of capacity is $(a + b)$.

The details of the entire process, as shown in Fig. 1, comprising the rework process, one-batch concept, and preempt concept, are described below.

1. The WIPs in a perfect production line is sent to node 1 via $a_0$, to node 2 via arc $a_1$, and output via arc $a_2$ (see the solid line in Fig. 1).

2. Some of the perfect WIPs in the production line become rework WIPs in the rework production line for the rework process (see the dashed line from nodes 2 to 1 in Fig. 1). The remaining perfect WIPs wait in node 2 until the rework WIPs return to node 2, and such a policy is called as the one-batch policy. Note that some of the defect WIPs are discarded without rework.

3. Some of the fixed rework WIPs are sent to node 1 via arc $a_3$ and to node 2 again via $a_4$ in the rework production line (see the dashed arc $a_4$ in Fig. 1). Note that these fixed rework WIPs can be defected and are discarded if they become defected again.

4. All the perfect (see the solid arc $a_2$) and rework (see the dashed arc $a_5$) WIPs exclude the discarded WIPs output from node 2. Note that each rework WIP consumes the same amount of capacity as each perfect WIP before it is sent to the rework process, i.e., via $a_1$, and utilizes extra capacity in the rework process via $a_4$ to return to node 2.

Rework networks already exist in many real-life applications in industrial manufacturing, as verified in [22-28]. Despite the discussion and correction of the errors in rework network reliability problems in [23, 27, 28], they still have many limitations in the current related literature. To establish a more general rework network with a more efficient algorithm to calculate its reliability, this study has two objectives:

1. introduce a new deterioration-effect multi-state multi-rework network based on the one-batch policy, preempt strategy, and deterioration effect with more than one rework process to achieve real-life applications.
2. propose a novel binary addition tree (BAT)-based algorithm to calculate the reliability of the proposed new rework network.

The remainder of the paper is organized as follows. The required notations and assumptions of the proposed multi-rework network are presented in Section 2. A summary of the existing rework networks and BATs is provided in Section 3. Section 4 discusses how to separate the entire new multi-rework network problem into subproblems based on WIPs and split nodes to reduce its complexity. In Section 4, a new multi-state BAT is proposed to search for feasible deterioration-effect multi-state vectors for the subproblems. In Section 5, another BAT called the nested multi-state BAT is proposed to combine the deterioration-effect multi-state vectors into feasible solutions for the proposed problem, together with a discussion of the reliability calculation. Section 6 presents the complete proposed algorithm, a demonstrated example, and the analysis of the proposed algorithm performance using five rework problems. Finally, the conclusion is presented in Section 7.

## 2. NOTATIONS, NOMENCLATURES, AND ASSUMPTIONS

The required notations, nomenclatures, and assumptions for the proposed one-batch preempt deterioration-effect multi-state multi-rework network reliability problem and the proposed algorithms are described in this section.

## 2.1 NOTATIONS

$m$ : number of arcs

$n$ : number of nodes

$|\bullet|$ : number of elements in $\bullet$

$V$ : node set $V = \{1, 2, \ldots, n\}$ and $|V| = n$

$e_{i,j}$ : directed arc from nodes $i$ to $j$, where nodes $i$ and $j$ are in $V$

$a_k$ : $k$th arc

$E$ : arc set $E = \{e_{i,j} = a_k \mid i, j \in V\}$ and $|E| = m$

$\mathbf{D}_{max}(i)$ : maximal capacity of node $i$ in $V$.

$\mathbf{D}_{min}(i)$ : minimal capacity of node $i$ in $V$.

$\mathbf{D}_{max}(a_{i,j})$ : maximal capacity of arc $a_{i,j}$ in $E$ and $\mathbf{D}_{max}(a_{i,j}) = \mathbf{D}_{min}(i)$.

$\mathbf{D}_{min}(a_{i,j})$ : minimal capacity of arc $a_{i,j}$ in $E$ and $\mathbf{D}_{min}(a_{i,j}) = \mathbf{D}_{min}(i)$.

$\mathbf{D}(i, l)$ : occurrent probability of node $i$ with capacity $l$.

$\mathbf{D}(a)$ : perfect rate of arc $a$ in $E$.

$\mathbf{D}$ : state distribution defining $\mathbf{D}_{max}(i)$, $\mathbf{D}_{min}(i)$, $\mathbf{D}(i, l)$, and $\mathbf{D}(a)$ for all $i \in V$ with capacity $l$ and $a \in E$.

$G(V, E, \mathbf{D})$ : one-batch preempt deterioration-effect multi-state multi-rework network with $V = \{1, 2, \ldots, n\}$, $E = \{a_k = e_{i,j} \mid i, j \in V\}$, state distribution $\mathbf{D}$, nodes 1 and $n$ are the source node and the sink node, respectively.

$\Pr(\bullet)$ : occurrent probability of $\bullet$

$b$ : number of input WIPs

$d$ : number of required output WIPs

$R_{b,d}$ : reliability such that at least $d$ units of output (including perfect and rework output) do not have any defect after inputting $b$ units into system

$F_f$ : $f$th product line

$\phi$ : number of product lines

$x_k$ : the number of WIPs in arc $a_k \in E$.

$x_{[k]}$ : the number of WIPs in arc $a_k \in E$.

$X(a_k)$ : the number of WIPs in arc $a_k \in E$ based on the state vector $X$.

$X$ : $X = (x_1, x_2, \ldots, x_m)$ is the vector represented the number of WIPs in each arc, where $X(a_k) = x_k$ for all $a_k \in E$.

$a_{[k]}$ : arc before $a_k$ in a product line

$x_{[k]}$ : number of WIPs in $a_{[k]}$

$z_k$ : $k$th vector (solution) obtained from the $k$th BAT

$z_{k,h}$ : value of the $h$th coordinate in the $k$th vector (solution) obtained from the $k$th BAT

$Z$ : $Z = (z_1, z_2, \ldots, z_\phi)$ is a vector (solution) obtained from the proposed nested multi-state BAT

$T$ : the runtime of the proposed algorithm

$m_i$ : the number of arcs in the $i$th product lines

$S_{b,d}$ : the total number of vectors (solutions) obtained the proposed nested multi-state BAT for input = $b$ and output = $d$

$s_{b,d}$ : the total number of feasible vectors (solutions) obtained the proposed nested multi-state BAT for input = $b$ and output = $d$

## 2.2 NOMENCLATURES

split node : A special node and one and only one rework process starts from each split node.

one-batch : All the perfect WIPs entered node $i$ need to wait for the rework WIPs from node $i$ to be reworked and then proceed to the next node together.

preempt : Each WIP consumes a resource in each arc and each node each time no matter how many times it has been reached into such arc or node.

| | | |
|---|---|---|
| multi-rework | : | A multi-rework network has many rework processes. |
| product line | : | A directed simple path and it is either a perfect product line a rework product line. There is only one perfect product line from the input to the output and each rework product line is from a split node to the output. |
| deterioration effect | : | The deterioration effect causes the number of WIPs is decreased or the same in each product line. |

## 2.3 ASSUMPTIONS

1) The number of WIPs is a non-negative integer-valued variable in relation to a given distribution.

2) The number of WIPs in different components are statistically independent.

3) All WIPs are not required to satisfy the conservation law in the network.

4) All WIPs obey the one-batch and preempt concepts.

5) Each rework WIP is discarded if it is failed again.

## 3. SUMMARY OF EXISTING REWORK NETWORKS AND BATs

The proposed rework problem is an extension of the "one-batch preempt single rework" which clarifies how perfect and rework WIPs share the capacities of each node. The proposed algorithm for the multi-rework problems is based on the BAT proposed by Yeh in [4, 5, 29, 30?]. Hence, the one-batch preempt single rework, multi-vector solution structure, and BAT are reviewed in this section.

### 3.1 One-Batch Policy, Preempt Strategy, and Single Rework

A single-rework network is a distinct rework network with only one rework process; an example is shown in Fig. 1. To verify whether a WIP is good before it is assembled in a device, a company has to ensure that the detection line is typically occupied for a certain working hour. All the WIPs, including the perfect and rework WIPs, need to enter and leave the production line together, and this concept is called as the one-batch policy here. For example, in Fig. 1, each perfect WIP processed from nodes 1 to 2 via $a_1$ needs to wait for the finishing of all associated rework WIPs in the same batch sharing the same nodes and arcs.

In a preempt strategy, a WIP still occupies the capacity of the related workstation provided there exists another WIP that needs to be reworked in the arc. In a preempt network, the capacity of each workstation is calculated by the summation of the number of perfect and rework WIPs that are processed in that workstation. Hence, the total capacity of each arc shared and used in different product lines must be less than or equal to its maximum capacity, i.e., the capacities of each node are shared by both normal and reworked products.

For example, in Fig. 1, let the perfect production line, $F_1 = \{a_0, a_1, a_2\}$, be for perfect WIPs, and the rework production line, $F_2 = \{a_3, a_4, a_5\}$, be for rework WIPs. Because the perfect and rework WIPs share the same capacity in the same batch in $a_1$, we have $x_1 + x_4 \leq W(a_1)$.

**3.2 Traditional BAT**

The traditional BAT was first proposed by Yeh in [5]. It can identify all the binary-state vectors from vector zero to vector one in a manner that is similar to adding one to vector zero repeatedly until it becomes vector one. For example, let the number of coordinates, $m = 3$, i.e., $X = (x_1, x_2, x_3)$. Accordingly, the binary representation of vector zero (0, 0, 0) is 000, which becomes 001 after adding one, i.e., the next vector to be found immediately after (0, 0, 0) is (0, 0, 1). Similarly, a BAT can find all the vectors easily, as can be seen from Table 2.

**Table 2.** The results obtained from BAT.

| $i$ | $x_1$ | $x_2$ | $x_3$ |
|---|---|---|---|
| 1 | 0 | 0 | 0 |
| 2 | 0 | 0 | 1 |
| 3 | 0 | 1 | 0 |
| 4 | 0 | 1 | 1 |
| 5 | 1 | 0 | 0 |
| 6 | 1 | 0 | 1 |
| 7 | 1 | 1 | 0 |
| 8 | 1 | 1 | 1 |

The pseudo-code of the traditional BAT is very simple and can be listed below:

ALGORITHM BAT

**STEP 0.** Let $i = m$ and $x_k = 0$ for $k = 0, 1, 2, \ldots, m$.

**STEP 1.** If $x_i = 0$, let $x_i = 1$, $i = m$, $X$ is a new vector, and go to STEP 1.

**STEP 2.** If $i = 0$, halt and all vectors are found.

**STEP 3.** Let $x_i = 0$, $i = i - 1$ and go to STEP 1.

BAT is an efficient and flexible method to find solutions such as the binary-state networks [5], the multi-state information networks [4], the propagation of wildfires [30], and the computer virus spread [34]. It can also be applied to other large-scale network issues in the future including the spread of computer viruses [31], infectious diseases [32], and social media [33].

### 3.3 Multi-state BAT

The traditional BAT identifies all the binary-state vectors from vector zero to vector one, as discussed in Section 3.2. However, in practical industrial production lines, the states of WIPs are not limited to 0 and 1, i.e., binary states only, and are multi states. Hence, the traditional BAT is extended to a multi-state BAT such that the maximum number of used states is not one only [5, 30]. The multi-state BAT proposed in [4] is adopted here to generate all the multi-state vectors to achieve the practical production lines in a rework process as follows.

ALGORITHM Multi-State BAT

**STEP 0.** Let $i = m$ and $x_k = 0$ for $k = 0, 1, 2, \ldots, m$.

**STEP 1.** If $x_i < \mathbf{D}_{\max}(a_i)$, let $x_i = x_i + 1$, $i = m$, $X$ is a new vector, and go to STEP 1.

**STEP 2.** If $i = 0$, halt and all vectors are found.

**STEP 3.** Let $x_i = 0$, $i = i - 1$ and go to STEP 1.

For example, let $m = 3$ and $\mathbf{D}_{max}(a_i) = 3$ for $i = 1, 2, 3$. All multi-state vectors generated in terms of the multi-state BAT are listed in Table 3.

Table 3. The results obtained from Multi-State BAT.

| $i$ | $x_1$ | $x_2$ | $x_3$ | $i$ | $x_1$ | $x_2$ | $x_3$ |
|---|---|---|---|---|---|---|---|
| 1 | 0 | 0 | 0 | 15 | 1 | 1 | 2 |
| 2 | 0 | 0 | 1 | 16 | 1 | 2 | 0 |
| 3 | 0 | 0 | 2 | 17 | 1 | 2 | 1 |
| 4 | 0 | 1 | 0 | 18 | 1 | 2 | 2 |
| 5 | 0 | 1 | 1 | 19 | 2 | 0 | 0 |
| 6 | 0 | 1 | 2 | 20 | 2 | 0 | 1 |
| 7 | 0 | 2 | 0 | 21 | 2 | 0 | 2 |
| 8 | 0 | 2 | 1 | 22 | 2 | 1 | 0 |
| 9 | 0 | 2 | 2 | 23 | 2 | 1 | 1 |
| 10 | 1 | 0 | 0 | 24 | 2 | 1 | 2 |
| 11 | 1 | 0 | 1 | 25 | 2 | 2 | 0 |
| 12 | 1 | 0 | 2 | 26 | 2 | 2 | 1 |
| 13 | 1 | 1 | 0 | 27 | 2 | 2 | 2 |
| 14 | 1 | 1 | 1 | | | | |

## 4. PROBLEM DIVISION

To solve the proposed problem efficiently, the proposed problem is divided into serval subproblems by splitting a production line into different production lines by splitting nodes. The solution of each production line is obtained using its own BAT and is discussed in this section. To yield real feasible solutions for the problem, these vectors obtained using different BATs must be combined to identify complete solutions. The details of the method to combine the solutions are provided in the next section.

### 4.1 Activity-on-node and Activity-on-arc Networks

In a one-batch preempt multi-state multi-rework network, each arc is perfectly reliable, and only the nodes have reliability problems. For example, in the state distribution, $\mathbf{D}$, as summarized in Table 1, of Fig. 1, only the nodes have states and their corresponding occurrent probabilities. Hence, the proposed one-batch preempt multi-state multi-rework network is originally an activity-on-node network (AON).

Based on Fig. 2, two different production lines are output from node 2, and it is impossible to obtain the value of each production line if the total production line of each node is represented by one variable only. To overcome the above problem in determining the values of rework and perfect production lines, the first step of the proposed algorithm for the problem is to transfer the AON to a AON with a distinct activity-on-arc network (AONA).

For example, there are two nodes in Fig. 1; after transferring the AON to an AONA, there are six arcs in Fig. 1, i.e., the number of coordinates in the multi-state vectors is increased from two to five. Note that the number of WIPs in the first arc $a_0$ is always equal to $b$ and $x_0 = b$ can be ignored in BAT for each rework network. Subsequently, the production line values can be easily determined from the coordinates.

To easily identify the production line a variable belongs to, the first subscript in the representation of these arcs, e.g., $a_{2,3}$ and $a_{1,4}$, is the production line label, and the second subscript labels the arc in the production line. For example, $a_{2,3}$ denotes the third arc in production line 2. Similarly, three benchmark reworks used in [23, 27, 28] are changed from the AON to an AOA, as shown in Fig.3 – Fig.5.

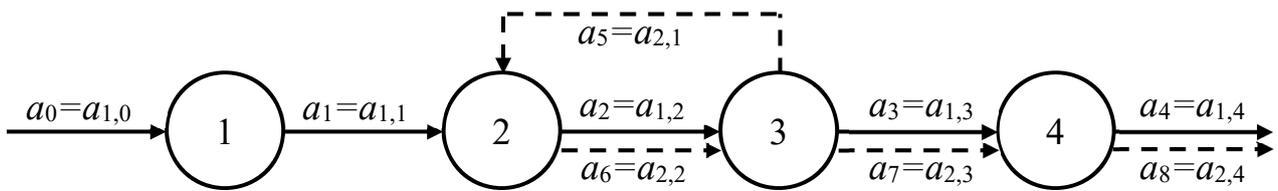

**Figure 3.** Benchmark rework problem.

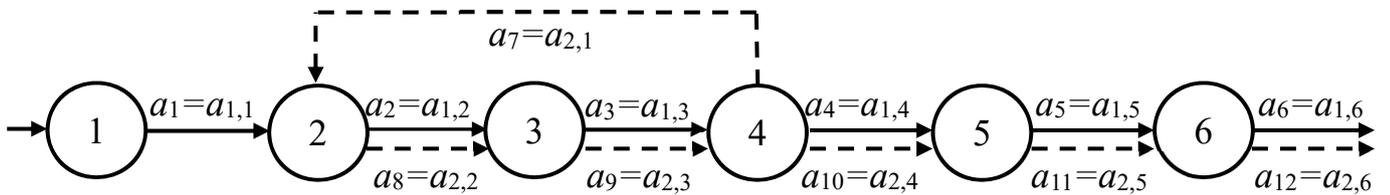

**Figure 4.** Benchmark rework problem.

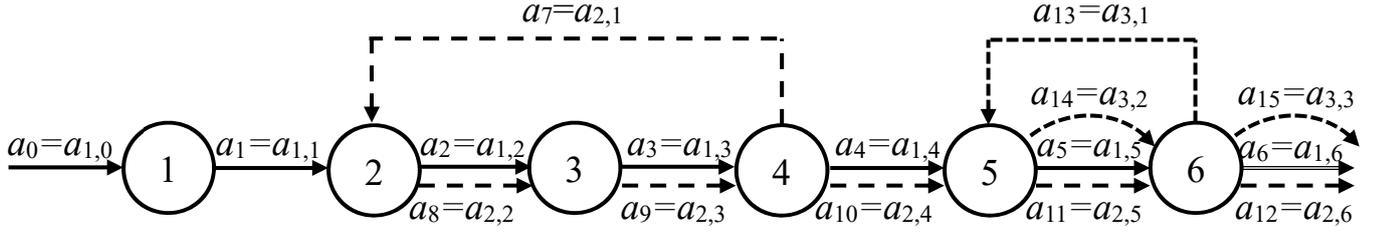

**Figure 5.** Benchmark rework problem.

### 4.2 Problem Division into Subproblems

To reduce the number of infeasible multi-state vectors, the proposed rework problem is divided into subproblems based on the production lines, including rework and perfect subproblems. Each rework is from a split node which is special nodes, e.g., node 2 in Fig. 1, and a split node can only be the start of one rework process. The basic concept is to find the production line from the input to the output such that there is one perfect production line. Moreover, all the rework production lines from the split nodes, e.g., node 2 in Fig. 1, start to yield output in the form of a rework subproblem. Hence, there is one and only one subproblem for each arc and no arc is shared in any two different subproblems, i.e., the solutions obtained from different subproblems are disjoint.

Without the loss of generality, both source node 1 and sink node $n$ are also the split nodes, e.g., node 1 is also a split node in Fig. 1. For example, the product line is split into perfect product line from input to node 1, from node 1 to 2, and from node 2 to the end (see the solid line), and the rework product line from nodes 2 to 1 to 2 (see the dashed line) in Fig. 1.

As discussed in Section 4.1, to easily identify different production lines, all the production lines are denoted as $F$, e.g., $F_1 = \{a_0, a_1, a_2\}$ is a perfect production line and $F_2 = \{a_3, a_4, a_5\}$ is a rework production line. Moreover, the $j$th arc in product line $F_i$ is denoted as $a_{i,j}$, e.g., $a_1 = a_{1,1}$ and $a_4 = a_{2,2}$ are the first and second arcs in $F_1$ and $F_2$, respectively.

Similarly, we obtain all the production lines for the rework networks shown in Fig. 1 and Fig.3 – Fig.5 as listed in the second column of Table 4.

Table 4. Production lines of benchmark rework networks.

| Networks | Product Lines | Node Constraints | Split Constraints | Deterioration effect |
|---|---|---|---|---|
| Fig. 1 | $F_1 = \{a_0, a_1, a_2\}$<br>$F_2 = \{a_3, a_4, a_5\}$ | $d \leq x_1 + x_4 \leq \text{Min}\{b, \mathbf{D}_{max}(1)\}$<br>$d \leq x_2 + x_3 + x_5 \leq \text{Min}\{b, \mathbf{D}_{max}(2)\}$ | $x_2 + x_3 \leq x_1$ | $x_2 \leq x_1 \leq x_0$<br>$x_5 \leq x_4 \leq x_3$ |
| Fig. 3 | $F_1=\{a_0, a_1, a_2, a_3, a_4\}$<br>$F_2=\{a_5, a_6, a_7, a_8\}$ | $d \leq x_1 \leq \text{Min}\{b, \mathbf{D}_{max}(1)\}$<br>$d \leq x_2 + x_6 \leq \text{Min}\{b, \mathbf{D}_{max}(2)\}$<br>$d \leq x_3 + x_5 + x_7 \leq \text{Min}\{b, \mathbf{D}_{max}(3)\}$<br>$d \leq x_4 + x_8 \leq \text{Min}\{b, \mathbf{D}_{max}(4)\}$ | $x_3 + x_5 \leq x_2$ | $x_{i+1} \leq x_i$<br>for all $i \neq 4$ |
| Fig. 4 | $F_1=\{a_0, a_1, a_2, a_3, a_4, a_5, a_6\}$<br>$F_2=\{a_7, a_8, a_9, a_{10}, a_{11}, a_{12}\}$ | $d \leq x_1 \leq \text{Min}\{b, \mathbf{D}_{max}(1)\}$<br>$d \leq x_2 + x_8 \leq \text{Min}\{b, \mathbf{D}_{max}(2)\}$<br>$d \leq x_3 + x_9 \leq \text{Min}\{b, \mathbf{D}_{max}(3)\}$<br>$d \leq x_4 + x_7 + x_{10} \leq \text{Min}\{b, \mathbf{D}_{max}(4)\}$<br>$d \leq x_5 + x_{11} \leq \text{Min}\{b, \mathbf{D}_{max}(5)\}$<br>$d \leq x_6 + x_{12} \leq \text{Min}\{b, \mathbf{D}_{max}(6)\}$ | $x_4 + x_7 \leq x_3$ | $x_{i+1} \leq x_i$<br>for all $i \neq 6$ |
| Fig. 5 | $F_1=\{a_0, a_1, a_2, a_3, a_4, a_5, a_6\}$<br>$F_2=\{a_7, a_8, a_9, a_{10}, a_{11}, a_{12}\}$<br>$F_3=\{a_{13}, a_{14}, a_{15}\}$ | $d \leq x_1 \leq \text{Min}\{b, \mathbf{D}_{max}(1)\}$<br>$d \leq x_2 + x_8 \leq \text{Min}\{b, \mathbf{D}_{max}(2)\}$<br>$d \leq x_3 + x_9 \leq \text{Min}\{b, \mathbf{D}_{max}(3)\}$<br>$d \leq x_4 + x_7 + x_{10} \leq \text{Min}\{b, \mathbf{D}_{max}(4)\}$<br>$d \leq x_5 + x_{11} + x_{14} \leq \text{Min}\{b, \mathbf{D}_{max}(5)\}$<br>$d \leq x_6 + x_{12} + x_{15} \leq \text{Min}\{b, \mathbf{D}_{max}(6)\}$ | $x_4 + x_7 \leq x_3$<br>$x_6 + x_{13} \leq x_5$ | $x_{i+1} \leq x_i$<br>for $i \neq 6, 12$ |

### 4.3 New Multi-state BAT for Subproblems

Each subproblem formed by a production line has its own solutions, which are obtained from the proposed BAT. Each solution of a subproblem is called a vector, and the solutions of the proposed problem are still referred to as solutions in the remainder of the paper to easily indicate where they are obtained.

In the traditional BAT or a multi-state BAT, the values of the coordinates in each vector decrease or increase based on the sequence order it generates. For example, based on Tables 2 and 3, the coordinates of $x_2$ can be less than, greater than, and/or equal to those of $x_1$ and/or $x_3$. However, the number of both the perfect and rework WIPs decreases in successive nodes owing to the defects. For example, assume the number of WIPs in node 1 is 14. Thus, it should be less than 14 in node 2 if there is at least one defective product. The above real-life phenomenon is called as the deterioration effect, which decreases the number of products occasionally [13-17], e.g., the last column in Table 4.

Regardless of using the traditional BAT or a multi-state BAT, all the found vectors are in the sequence from vector zero to the largest vector, e.g., (0, 0, 0), (0, 0, 1), …, (2, 2, 1), and (2, 2, 2), as listed in Table 3. In some scenarios, e.g., the proposed rework problem, all the vectors must be in the sequence from the largest to vector zero, e.g., (2, 2, 2), (2, 2, 1), …, (0, 0, 1), and (0, 0, 0).

To include the above two scenarios, a multi-state BAT must be revised such that the value of $x_{i+1}$ is not larger than that of $x_k$ for $k = 1, 2, …, i–1$ in each obtained multi-state vector. Hence, a top-down deterioration-effect multi-state BAT is proposed, and the details of its pseudo code are provided below.

ALGORITHM Top-Down Deterioration-Effect Multi-State BAT

**STEP 0.** $X = (D_{max}(a_1), D_{max}(a_2), …, D_{max}(a_m))$ is a feasible multi-state vector and let $i = m$.

**STEP 1.** If $x_i > 0$, let $x_i = x_i – 1$, $x_j = x_i$ for all $j > i$, $i = m$, $X$ is a feasible multi-state vector, and go to STEP 1.

**STEP 2.** If $i = 1$, halt and all vectors are found.

**STEP 3.** Let $i = i – 1$ and go to STEP 1.

For example, in Fig. 1 and product lines listed in Table 4, all feasible deterioration-effect multi-state vectors for $F_1$ are listed in Table 5.

Table 5. The feasible multi-state vectors for $F_1$.

| $i$ | $x_1$ | $x_2$ |
|---|---|---|
| 0 | 5 | 5 |
| 1 | 5 | 4 |
| 2 | 5 | 3 |
| 3 | 5 | 2 |
| 4 | 5 | 1 |
| 5 | 5 | 0 |
| 6 | 4 | 4 |
| 7 | 4 | 3 |
| 8 | 4 | 2 |
| 9 | 4 | 1 |
| 10 | 4 | 0 |
| 11 | 3 | 3 |
| 12 | 3 | 2 |
| 13 | 3 | 1 |
| 14 | 3 | 0 |

The proposed top-down deterioration-effect multi-state BAT is implemented to determine all the feasible deterioration-effect multi-state vectors for $F_2$, which is shown in Fig. 1, as follows.

**Solution:**

**STEP 0.** $X = (5, 5, 5)$ is the first feasible multi-state vector and let $i = 3$.

**STEP 1.** Because $x_3 = 5$, let $x_3 = x_3 - 1 = 4$, and go to STEP 1. Note that $X = (5, 5, 4)$ is a new vector.

:

**STEP 1.** Because $x_3 = 0$, go to STEP 2.

**STEP 2.** Because $i = 3 \neq 1$, go to STEP 3.

**STEP 3.** Let $i = i - 1 = 2$ and go to STEP 1.

**STEP 1.** Because $x_2 = 5$, let $x_2 = x_2 - 1 = 4$, $x_3 = x_2 = 4$, and $i = 3$, i.e., $X = (5, 4, 4)$ is a new vector, and go to STEP 1.

**STEP 1.** Because $x_3 = 4$, let $x_3 = x_3 - 1 = 3$, i.e., $X = (5, 4, 3)$ is a new vector, and go to STEP 1.

:

**STEP 1.** Because $x_2 = 0$, go to STEP 2.

**STEP 2.** Because $i = 2 \neq 1$, go to STEP 3.

**STEP 3.** Let $i = i - 1 = 1$ and go to STEP 1.

**STEP 1.** Because $x_1 = 5$, let $x_1 = x_1 - 1 = 4$ and $x_3 = x_2 = x_1 = 4$, i.e., $X = (4, 4, 4)$ is a new vector, and go to STEP 1.

**STEP 1.** Because $x_3 = 4$, let $x_3 = x_3 - 1 = 3$, i.e., $X = (4, 4, 3)$ is a new vector, and go to STEP 1.

:

**STEP 1.** Because $x_1 = 1$, let $x_1 = x_1 - 1 = 0$ and $x_3 = x_2 = x_1 = 0$, i.e., $X = (0, 0, 0)$ is a new vector, and go to STEP 1.

**STEP 2.** Because $i = 1$, halt and all feasible multi-state vectors are found.

The results of the deterioration-effect multi-state vectors obtained by the proposed top-down deterioration-effect multi-state BAT are provided in Table 6.

Table 6. Deterioration-effect multi-state vectors for $F_2$.

| $i$ | $x_3$ | $x_4$ | $x_5$ | $i$ | $x_3$ | $x_4$ | $x_5$ | $i$ | $x_3$ | $x_4$ | $x_5$ |
|---|---|---|---|---|---|---|---|---|---|---|---|
| 0 | 5 | 5 | 5 | 21 | 4 | 4 | 4 | 36 | 3 | 3 | 3 |
| 1 | 5 | 5 | 4 | 22 | 4 | 4 | 3 | 37 | 3 | 3 | 2 |
| 2 | 5 | 5 | 3 | 23 | 4 | 4 | 2 | 38 | 3 | 3 | 1 |
| 3 | 5 | 5 | 2 | 24 | 4 | 4 | 1 | 39 | 3 | 3 | 0 |
| 4 | 5 | 5 | 1 | 25 | 4 | 4 | 0 | 40 | 3 | 2 | 2 |
| 5 | 5 | 5 | 0 | 26 | 4 | 3 | 3 | 41 | 3 | 2 | 1 |
| 6 | 5 | 4 | 4 | 27 | 4 | 3 | 2 | 42 | 3 | 2 | 0 |
| 7 | 5 | 4 | 3 | 28 | 4 | 3 | 1 | 43 | 3 | 1 | 1 |
| 8 | 5 | 4 | 2 | 29 | 4 | 3 | 0 | 44 | 3 | 1 | 0 |
| 9 | 5 | 4 | 1 | 30 | 4 | 2 | 2 | 45 | 3 | 0 | 0 |
| 10 | 5 | 4 | 0 | 31 | 4 | 2 | 1 | 46 | 2 | 2 | 2 |
| 11 | 5 | 3 | 3 | 32 | 4 | 2 | 0 | 47 | 2 | 2 | 1 |
| 12 | 5 | 3 | 2 | 33 | 4 | 1 | 1 | 48 | 2 | 2 | 0 |
| 13 | 5 | 3 | 1 | 34 | 4 | 1 | 0 | 49 | 2 | 1 | 1 |
| 14 | 5 | 3 | 0 | 35 | 4 | 0 | 0 | 50 | 2 | 1 | 0 |
| 15 | 5 | 2 | 2 | | | | | 51 | 2 | 0 | 0 |
| 16 | 5 | 2 | 1 | | | | | 52 | 1 | 1 | 1 |
| 17 | 5 | 2 | 0 | | | | | 53 | 1 | 1 | 0 |
| 18 | 5 | 1 | 1 | | | | | 54 | 1 | 0 | 0 |
| 19 | 5 | 1 | 0 | | | | | 55 | 0 | 0 | 0 |
| 20 | 5 | 0 | 0 | | | | | | | | |

Based on Table 6, 56 multi-state vectors are obtained using the proposed deterioration-effect multi-state BAT, and the number of vectors is far less than that obtained by the traditional multi-state BAT, which is $6^3 = 216$. Hence, the proposed top-down deterioration-effect multi-state BAT can assist in finding all the deterioration-effect multi-state vectors without consuming time to filter these deterioration-effect multi-state vectors from all the multi-state vectors.

## 5. COMBINING VECTORS TO FORM SOLUTIONS

To determine the feasible solutions representing the feasible production line values in the proposed rework network, the deterioration-effect multi-state vectors generated from the disjoint subproblems must be combined into feasible solutions. The details of the method are provided in this section.

## 5.1 Basic Concept

Assume that there are $\phi$ production lines corresponding to $\phi$ disjoint top-down deterioration-effect multi-state BATs. Because each arc belongs to one and only one BAT, $\{X_1, X_2, \ldots, X_\phi\}$ is a possible multi-state vector for the problem, where $X_i$ is one of the multi-state vectors obtained from the $i^{th}$ BAT, where $i = 1, 2, \ldots, \phi$. For example, $(x_1, x_2) = (5, 4)$, and $(x_3, x_4, x_5) = (1, 1, 1)$ are one of the multi-vectors obtained from the first and second BATs, respectively, as listed in Tables 5 and 6. Hence, $(x_1, x_2, x_3, x_4, x_5) = (5, 4, 1, 1, 1)$ is one possible multi-state vector of the rework network shown in Fig. 1.

The above concept to combine vectors into solutions is very simple, and the number of possible feasible solutions is equal to the product of the numbers of multi-state vectors obtained from all the BATs, which include numerous infeasible multi-state vectors, i.e., low efficiency.

Hence, before combining all the vectors obtained from the subproblems into feasible solutions for the proposed problem, to reduce the runtime, we need to build the constraints to ensure each obtained solution is feasible and to discard the infeasible combined vectors before they form a complete solution. For example, let $X_{1,i} = (x_1, x_2)$ and $X_{2,j} = (x_4, x_5, x_6)$ be the feasible vectors obtained from $F_1$ and $F_2$ (of Fig. 1), respectively, as listed in Tables 5 and 6. If the combination of $X_{1,i}$ and $X_{2,j}$ is infeasible, then we can confirm that the combination of the two vectors is also feasible and can discard it without further testing.

## 5.2 Constraints

For combining the feasible multi-state vectors into feasible solutions with increased efficiency, two constraints are proposed: node and split constraints, e.g., the third and fourth columns of Table 4, respectively. These two constraints are related vectors that share the same arcs, and their details are discussed below.

**5.2.1 Node Constraints**

The total production line value of all the arcs from or to a particular node must not be larger than the capacity of this node owing to the preempt characteristic of the rework network. Moreover, it should not be less than the required amount of production line $d$, i.e.,

$$\text{Max}\{d, \mathbf{D}_{\max}(k)\} \leq \sum_{\forall i \in I} x_i \leq \text{Min}\{b, \mathbf{D}_{\max}(k)\}, \tag{1}$$

where all $a_i$ in arc subset $I$ are from the same node $k$. Eq. (1) is called as the node (capacity) constraint and is one of two cores to combine the multi-state vectors generated from the disjoint top-down deterioration-effect multi-state BATs into feasible solutions of the proposed problem.

For example, let $b = 5$ be the total production line entering node 1, and $d = 3$ be the required production line based on Fig. 1. Owing to the defective effect, arcs $a_1$ and $a_4$ are from node 1, arcs $a_2$ and $a_5$ are from node 2, and the production line leaving any node is less than or equal to 5, i.e., $x_1 + x_4 \leq \text{Min}\{5, \mathbf{D}_{\max}(1)\} = \text{Min}\{5, 7\} = 5$ for node 1 and $x_2 + x_3 + x_5 \leq \text{Min}\{5, \mathbf{D}_{\max}(2)\} = \text{Min}\{5, 5\} = 5$ for node 2 in Fig. 1.

Also, all the production lines leaving any node that is not inside the rework process must be larger than or equal to $d = 3$, i.e., $\text{Max}\{d = 3, \mathbf{D}_{\min}(1)\} = \text{Max}\{3, 0\} = 3 \leq x_1 + x_4$ for node 1 and $\text{Max}\{d = 3, \mathbf{D}_{\min}(2)\} = 3 \leq x_2 + x_3 + x_5$ for node 2. Hence, we have two node constraints, as listed in the third column of Table 4 for Fig. 1.

Similarly, all the node limitations corresponding to Fig.3 – Fig.5 are listed in the third column of Table 4.

**5.2.2 Split Constraints**

The production line is split into a perfect production line, the rework production line, and discarded production line. The core to combining multi-state vectors feasibly is that the sum of the first two different production lines is equal to the production line before the split. The number of discarded production lines is not recorded. Hence, in the split nodes, the sum of the perfect and rework production lines is less than the original production line, i.e.,

$$\sum_{\forall i \in I} x_i \leq x_k, \tag{2}$$

where all WIPs in $a_i \in I$ are from $a_k$ via the same split node and the production line is immediately sent to rework. Eq. (2) is called as the split constraint. For example, node 2 is the only split node in Fig. 1, and

$$x_2 + x_3 \leq x_1 \tag{3}$$

because $x_1$ units of the production line in $a_2$ are split into $x_2$ and $x_3$ after node 2. Similarly, the split limitations corresponding to Fig.3 – Fig.5 are listed in the fourth column of Table 4.

For example, $(x_1, x_2, x_3, x_4, x_5) = (5, 4, 1, 1, 1)$ violates the node constraint, $x_2 + x_3 + x_5 \leq \text{Min}\{b = 5, \mathbf{D}_{\max}(2) = 5\} = 5$ and is an infeasible multi-state vector of Fig. 1.

## 5.3 Proposed New Nested Multi-state BAT for Combining Solutions

Another new multi-state BAT called as the nested multi-state BAT is proposed to combine all the vectors obtained from different BATs to form solutions for the proposed problem.

There is a one-to-one relationship between a production line and a BAT, as discussed in Section 4.2. Following considering node and split limitations, $\{X_1, X_2, \ldots, X_\phi\}$ is a feasible multi-state vector for the problem, where $X_i$ is one of the multi-state vectors obtained from the $i^{th}$ BAT for $i = 1, 2, \ldots, \phi$, if both the node and split limitations are satisfied.

The multi-state vector generated from the proposed nested BAT is a $\phi$-tube vector. Let $(x_1, x_2, \ldots, x_\phi)$ be one of these vectors, where $x_i = k$ be the $k$th multi-vector selected from the $i$th subproblem and $k = 1, 2, \ldots,$ (the number of solutions obtained from the $i$th BAT). For example, Fig. 1 displays two production lines, i.e., two subproblems and solution $X = (1, 52)$ denoting the first and the $52^{nd}$ multi-vectors of the first and second BATs corresponding to the first and second subproblems. Hence, according to Tables 5 and 6, the corresponding solution to $X = (1, 52)$ is $(5, 4, 1, 1, 1)$, where $(5, 4)$ and (1, 1, 1) are the first and the 52$^{nd}$ multi-vectors vectors obtained from the first and second BATs, respectively.

The pseudo-code of the proposed nested multi-state BAT for combining the vectors obtained based on the BAT corresponding to each subproblem is provided below.

ALGORITHM Constrained Nested Multi-state BAT

**STEP 0.** Let $i = \phi$, $Z = (z_1, z_2, \ldots, z_\phi)$ be vector zero, then go to STEP 3.

**STEP 1.** If $z_i < z_{i,UB}$, let $z_i = z_i + 1$ and $i = \phi$. Then go to STEP 3.

**STEP 2.** If $i = 1$, stop. Otherwise, let $z_i = 0$, $i = i - 1$, and go to STEP 1.

**STEP 3.** If solution $X$ corresponding to $Z = (z_1, z_2, \ldots, z_\phi)$ satisfies all the constraints, then $X$ is a feasible solution for the problem, and go to STEP 1.

For example, let $b = 5$ and $d = 3$ in Fig. 1. All the deterioration-effect multi-state vectors for two production lines $F_1$ and $F_2$ are listed in Tables 5 and 6, respectively. For ease of understanding, the proposed algorithm is demonstrated by an example shown in Fig. 1 to establish the general step-by-step procedure to calculate the one-batch preempt multi-state multi-rework network reliability.

**Solution:**

**STEP 0.** Let $i = \phi = 2$, $Z = (z_1, z_2) = (0, 0)$, and go to STEP 3.

**STEP 3.** Because the zeroth vectors in Tables 5-6 are $(x_1, x_2) = (5, 5)$ and $(x_3, x_4, x_5) = (5, 5, 5)$, respectively, $X = (x_1, x_2, x_3, x_4, x_5) = (5, 5, 5, 5, 5)$ is the solution corresponding to $Z = (0, 0)$. $X$ violates $3 \leq x_1 + x_4 \leq 5$, i.e., $X$ is infeasible, and go to STEP 1.

**STEP 1.** Because $z_2 = 0 < 6$, let $z_2 = z_2 + 1 = 1$, $i = 2$, and go to STEP 3.

**STEP 3.** Because $Z = (z_1, z_2) = (0, 1)$ and the zeroth and first vectors in Tables 5-6 are $(x_1, x_2) = (5, 5)$ and $(x_3, x_4, x_5) = (5, 5, 4)$, respectively, $X = (x_1, x_2, x_3, x_4, x_5) = (5, 5, 5, 5, 4)$ is the solution corresponding to $Z = (0, 1)$. $X$ also violates $3 \leq x_1 + x_4 \leq 5$, and go to STEP 1.

:

**STEP 3.** Because $Z = (z_1, z_2) = (2, 51)$ and the second and 51$^{st}$ vectors in Tables 5-6 are $(x_1, x_2) = (5, 3)$ and $(x_3, x_4, x_5) = (2, 0, 0)$, respectively, $X = (x_1, x_2, x_3, x_4, x_5) = (5, 3, 2, 0, 0)$ is the solution corresponding to $Z = (2, 51)$, $X$ is feasible, and go to STEP 1.

**STEP 1.** Because $z_2 = 51 < 55$, let $z_2 = z_2 + 1 = 52$, $i = 2$, and go to STEP 3.

**STEP 3.** Because $Z = (z_1, z_2) = (2, 52)$, the second and 52$^{nd}$ vectors in Tables 5-6 are $(x_1, x_2) = (5, 3)$ and $(x_3, x_4, x_5) = (1, 1, 1)$, respectively, $X = (x_1, x_2, x_3, x_4, x_5) = (5, 3, 1, 1, 1)$ are the solutions corresponding to $Z = (2, 52)$, $X$ violates $3 \leq x_1 + x_4 \leq 5$, and go to STEP 1.

:

There are 16 feasible solutions among the $15 \times 56 = 840$ solutions generated by the above BATs. The results of all the feasible vectors and solutions are listed in Table 7.

Table 7. The results of the BAT for Fig. 1.

| $i$ | $j$ | $Z_i$ | $X_j$ | $\Pr(X_j)$ |
|---|---|---|---|---|
| 55 | 1 | (0, 55) | (5, 5, 0, 0, 0) | 0.00828039319235627200 |
| 110 | 2 | (1, 54) | (5, 4, 1, 0, 0) | 0.00028751379294180514 |
| 111 | 3 | (1, 55) | (5, 4, 0, 0, 0) | 0.00044724358503033483 |
| 163 | 4 | (2, 51) | (5, 3, 2, 0, 0) | 0.00000399324907460150 |
| 166 | 5 | (2, 54) | (5, 3, 1, 0, 0) | 0.00000621171949272477 |
| 167 | 6 | (2, 55) | (5, 3, 0, 0, 0) | 0.00002129731899183067 |
| 391 | 7 | (6, 55) | (4, 4, 0, 0, 0) | 0.00002710563823667484 |
| 444 | 8 | (7, 52) | (4, 3, 1, 1, 1) | 0.00019312771114752183 |
| 445 | 9 | (7, 53) | (4, 3, 1, 1, 0) | 0.00000715287900611796 |
| 446 | 10 | (7, 54) | (4, 3, 1, 0, 0) | 0.00000075293476321669 |
| 447 | 11 | (7, 55) | (4, 3, 0, 0, 0) | 0.00000258149001271632 |
| 497 | 12 | (8, 49) | (4, 2, 2, 1, 1) | 0.00000402349594742319 |
| 500 | 13 | (8, 52) | (4, 2, 1, 1, 1) | 0.00000312938508700485 |
| 671 | 14 | (11, 55) | (3, 3, 0, 0, 0) | 0.00000004345937559720 |
| 724 | 15 | (12, 52) | (3, 2, 1, 1, 1) | 0.00000018965946725842 |
| 774 | 16 | (13, 46) | (3, 1, 2, 2, 2) | 0.00000033783099377012 |
| SUM | | | | 0.00928509734192486950 |

## 5.4 Calculation of Probability

To calculate the reliability of the rework network, we can simply add the probabilities of all the feasible solutions because these solutions are independent. The probability of each feasible solution is the product of the probabilities of the production lines in the arcs, which are based on the combinations and states of nodes according to the state distribution, **D**.

Let node [$i$] be the node immediately before node $i$ along production line $F_j$. Then the probability of the combination of $x_{[i-1]}$ units of the production lines input to node $i$ and $x_i$ units of the production line output from node $i$ can be calculated as follows:

$$\Pr(x_i, x_{[i]}) = \frac{x_{[i]}!}{x_i!\,(x_{[i]} - x_i)!}(1-\delta_i)^{x_i} \cdot \delta_i^{(x_{[i]}-x_i)}. \tag{4}$$

Furthermore, the probability to obtain feasible solution $X = (x_1, x_2, \ldots, x_m)$ is

$$\Pr(X) = \prod_{i,j}\Pr(x_i, x_{[i]}) \cdot \prod_{k=1}\mathbf{D}(k, l_k), \tag{5}$$

where for $a_i \in F_j$, $i = 1, 2, \ldots, m$ and $j = 1, 2, \ldots, \phi$.

For example, the 309$^{th}$ solution, (5, 3, 2, 0, 0), is also the first feasible solution, as provided in Table 7 for Fig. 1.

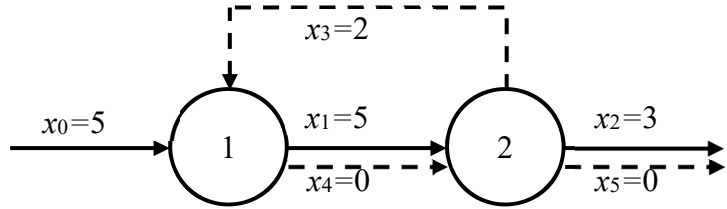

**Figure 6.** Rework for feasible solution $(x_1, x_2, x_3, x_4, x_5) = (5, 3, 2, 0, 0)$ in Fig. 1.

From Eq. (4), we have

$$\Pr(x_1 = 5,\ \text{input} = x_0 = b = 5) = \frac{b!}{x_1!\,(b-x_1)!}(1-\delta_{1,1})^{x_1} \cdot \delta_{1,1}^{b-x_1}$$

$$= \frac{5!}{5!\,0!}(0.99)^5 \cdot (0.01)^0$$

$$= 0.950990 \tag{6}$$

$$\Pr(x_2 = 3, x_1 = 5) = \frac{5!}{3!\,(5-2)!}(0.9)^3 \cdot (0.1)^2 = 0.072900 \tag{7}$$

$$\Pr(\text{output} = 3, x_2 = 3) = \frac{3!}{3!\,0!}(0.8)^3 \cdot (0.2)^0 = 0.512000 \tag{8}$$

$$\Pr(x_4 = 0, x_3 = 2) = \frac{2!}{2! \cdot 0!}(0.95)^0 \cdot (0.05)^2 = 0.002500 \tag{9}$$

$$\Pr(x_5 = 0, x_4 = 0) = \frac{0!}{0! \cdot 0!}(0.7)^0 \cdot (0.3)^0 = 1 \tag{10}$$

$$\Pr(\text{output} = 0, x_2 = 0) = \frac{0!}{0! \cdot 0!}(0.9)^0 \cdot (0.2)^0 = 1 \tag{11}$$

$$\mathbf{D}(1, x_1 + x_4 = 5) = 0.05 \tag{12}$$

$$\mathbf{D}(2, x_2 + x_3 + x_5 = 5) = 0.9 \tag{13}$$

Hence, the occurrent probability to have $X_{309} = (5, 3, 2, 0, 0)$ is

$$\prod_{i,j} \Pr(x_i, x_{[i]}) \cdot \prod_{k=1}^{2} \mathbf{D}(k, l_k)$$

$= 0.950990 \cdot 0.072900 \cdot 0.512000 \cdot 0.002500 \cdot 1 \cdot 1 \cdot 0.05 \cdot 0.09$

$$= 0.00000399324907460150. \tag{14}$$

Similarly, the probability of each solution, i.e., $\Pr(X_j)$ for $j = 1, 2, \ldots, 16$, is provided in the last column of Table 7.

## 6. PROPOSED NEW MULTI-STATE BATS

In this section, the proposed algorithm based on BATs, its validation, and testing on different rework problems that are discussed in the literature is presented.

### 6.1 Proposed Algorithm

The proposed algorithm has four stages for solving the one-batch preempt multi-state multi-rework network reliability problem, which are as follows:

1. The proposed BAT-based algorithm needs to separate the rework problem into subproblems based on the production line and the split nodes.

2. The proposed top-down deterioration-effect multi-state BAT is applied to each subproblem to obtain the related feasible deterioration-effect multi-state vectors.

3. The proposed nested multi-state BAT is applied to combine these feasible deterioration-effect multi-state vectors into feasible solutions for the proposed rework problem.

4. The occurrent probability of the feasible solutions is calculated, and the summation of these probabilities is the reliability of the one-batch preempt multi-state multi-rework network.

The pseudocode of the proposed algorithm based on the above concepts for the related rework problem is provided below.

**Input:** One-batch preempts multi-state multi-rework network $G(V, E, \mathbf{D})$ with source node 1, sink node $n$, input value $b$, and output value $d$.

**Output:** Reliability $R$ to input $b$ units of products and output at least $d$ units of products.

**STEP 0.** The AON is changed to an AOA, as discussed in Section 4.1.

**STEP 1.** The proposed rework network is separated into subproblems based on the perfect and rework production lines, e.g., $F_1, F_2, \ldots, F_\phi$.

**STEP 2.** All feasible deterioration-effect multi-state vectors of $F_i$ are found using the top-down deterioration-effect BAT proposed in Section 4.3 for all $i = 1, 2, \ldots, \phi$.

**STEP 3.** The proposed nested BAT is implemented together (proposed in Section 5.3) with the node constraints and the split constraints (proposed in Section 5.2) to combine all the feasible deterioration-effect multi-state vectors into feasible solutions.

**STEP 4.** The probability of each feasible solution is calculated using Eq. (5), and the reliability is obtained by the summation of all the probabilities.

The flowchart of the proposed algorithm is depicted in Fig. 7.

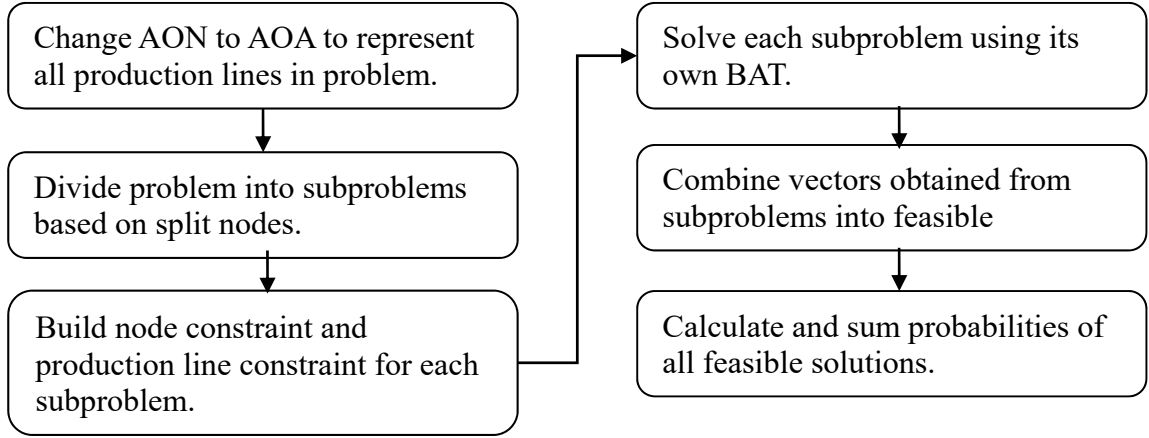

**Figure 7.** Flowchart of the proposed algorithm for the proposed Rework Problems.

**6.2 Demonstrated Example**

The rework network depicted in Fig. 1 with $b = 5$ and $d = 3$ is selected to illustrate the general procedures of the proposed algorithm for calculating the rework network reliability as follows:

**STEP 0.** The AON is transferred to an AOA, as shown in Fig. 1.

**STEP 1.** The problem is separated into subproblems based on the production lines and the split nodes, i.e., the dotted, dashed, and solid lines in Fig. 1.

**STEP 2.** All the feasible deterioration-effect multi-state vectors for $F_1$ and $F_2$ are found based on the proposed top-down deterioration-effect multi-state BAT, and the results are shown in Tables 5 and 6.

**STEP 3.** All the feasible deterioration-effect multi-state vectors obtained from $F_1$ and $F_2$ based on the proposed nested multi-state BAT are calculated, and the results are shown in Fig. 7.

**STEP 4.** The probabilities of all the feasible solutions, as listed in the last column of Table 7, are calculated and added to yield the reliability of the reliability of the rework network, which is depicted in Fig .1. Hence, the reliability is 0.00928509734192486950.

**6.3 Computational Experiments**

The rework network shown in Fig. 1 under the setting of **D** in Table 1 is named TEST 1. To further examine the performance of the proposed algorithm for calculating the reliability of the proposed new rework problem, four additional moderate-size rework networks, as shown in Fig.1, Fig.3 – Fig. 5 [23, 27, 28], are named TESTs 2-5, respectively, and tested under the state distribution **D** listed below:

$$\mathbf{D}(i, l) = 0.1 \text{ for all } i \in V \text{ and } l = 0, 1, \ldots, 9 \text{ and } \mathbf{D}(a) = 0.99 \text{ for all } a \in E. \tag{15}$$

The proposed algorithm is implemented in Dev C++ run on Windows 10 (64bit) with Intel(R) Core(TM) i7-8650U CPU @ 1.90GHz, 2.11 GHz and 16 GB of RAM. As listed in Table 4, there are 2, 2, and 3 top-down deterioration-effect multi-state BATs for the 2, 2, and 3 production lines in Fig.3 – Fig.5, respectively. The entire data of these rework networks are adopted from [23, 27, 28].

To verify the performance of the proposed BAT, the complete results obtained from the proposed algorithm for all five tests are listed in Appendix. All the number of feasible deterioration-effect multi-state sub-vectors corresponding to product lines of these rework problems are listed in Appendix under title $m_i$ for all the $i$th BAT corresponding to the $i$th product line. The proposed nested multi-state BAT is applied to combine these vectors to filter out feasible solutions to solve each rework problem and the number of all solutions $S_{b,d}$ and feasible solutions $s_{b,d}$ as listed in Appendix. The final reliability $R_{b,d}$ and the runtime T of each rework network are also listed in Appendix, respectively.

It is trivial that $d \leq b$ for all input $b$ and output $d$. Based on Appendix, we have Fig. 8 and the following general observations:

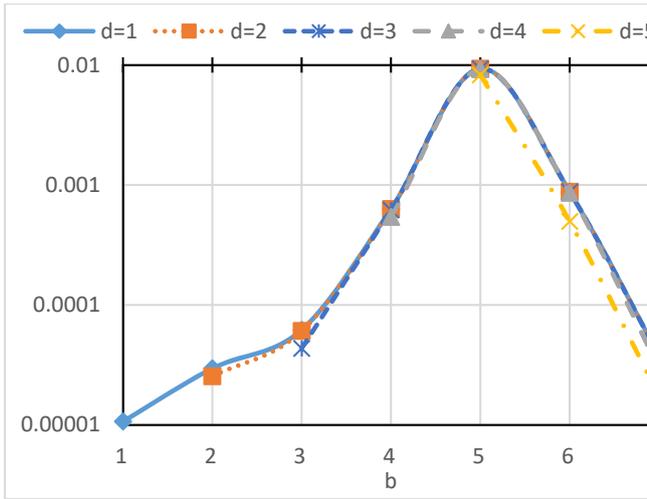
(a)
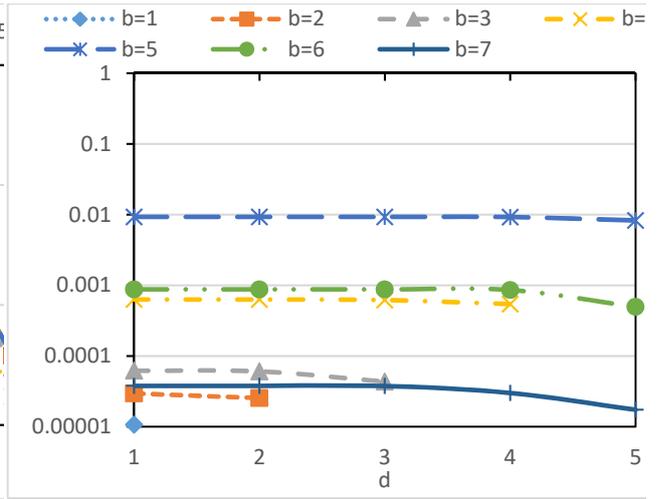
(b)
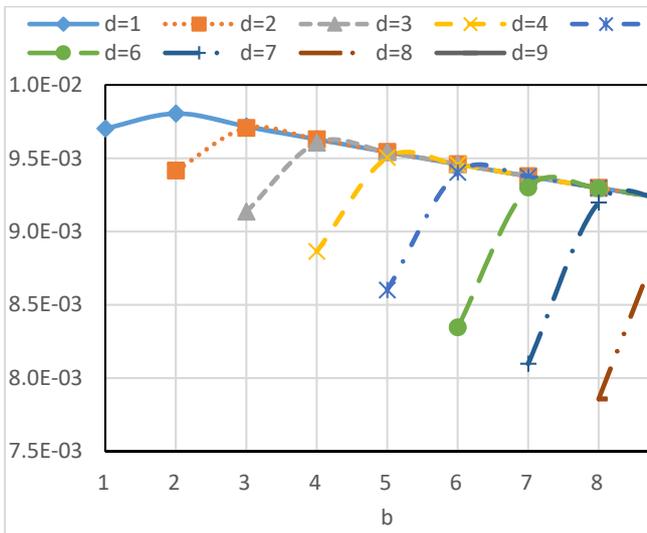
(c)
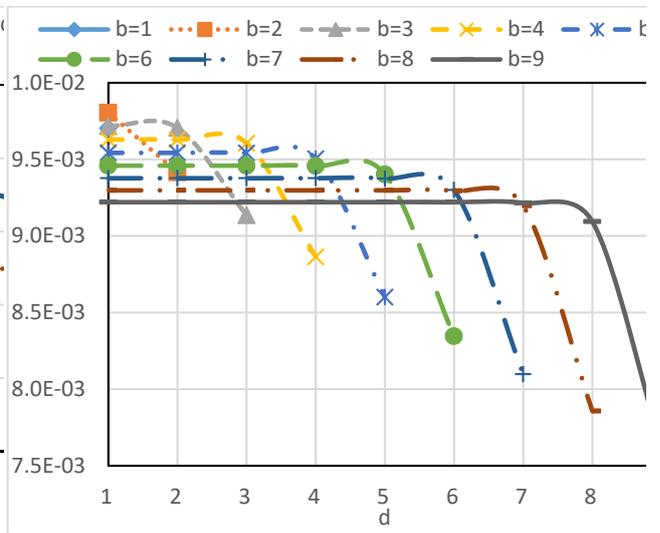
(d)
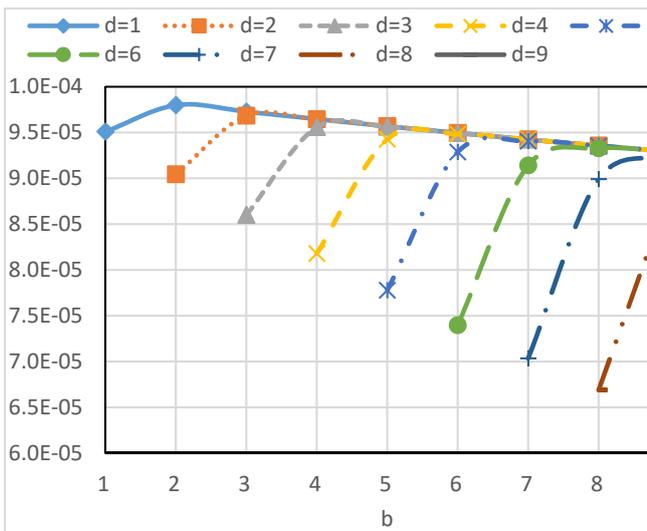
(e)
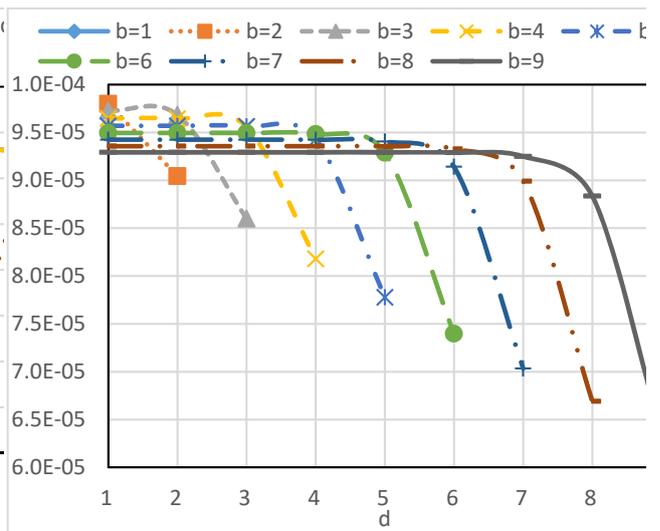
(f)

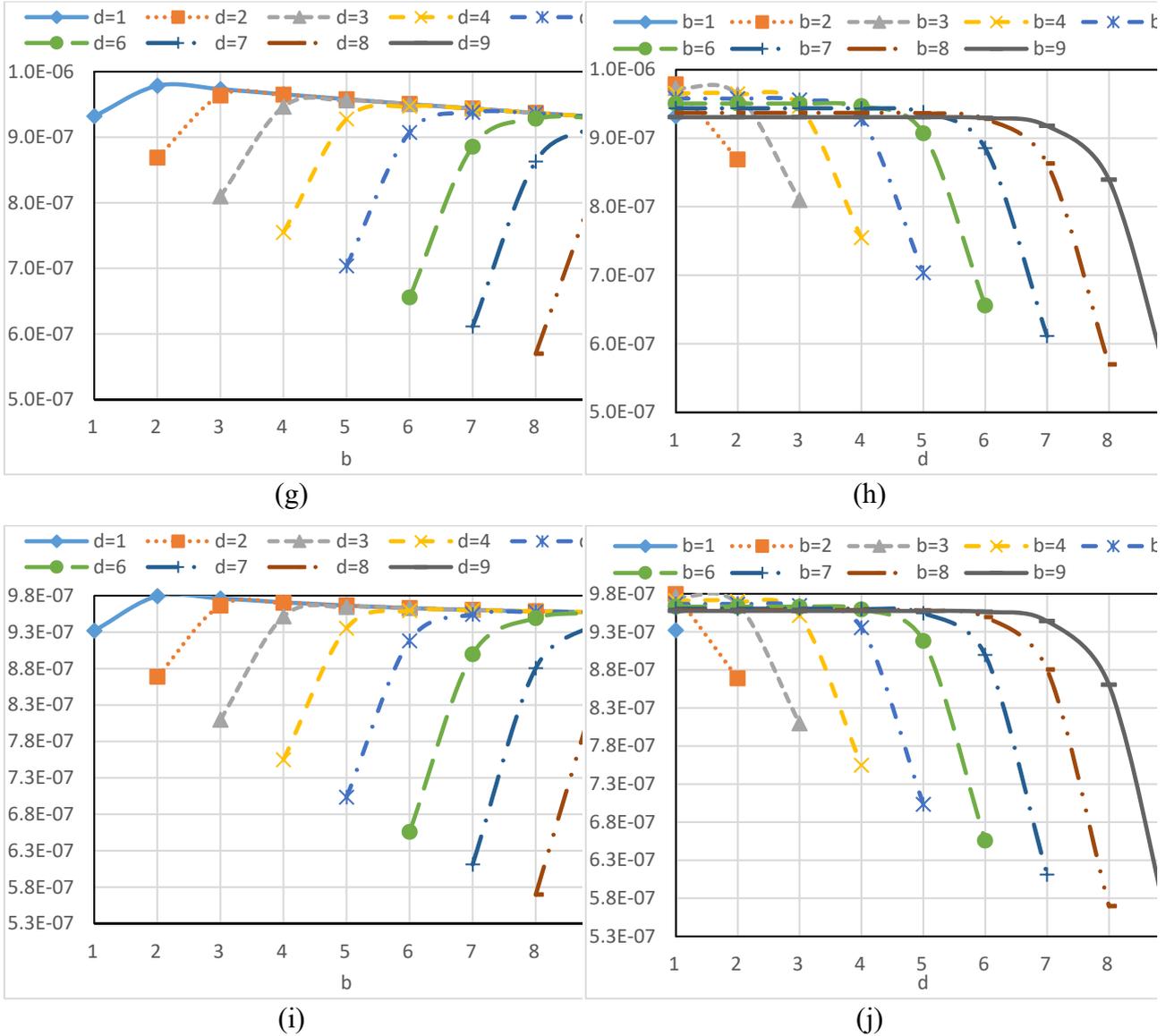

**Figure 8.** Results based on the values of $b$ and $d$ ($y$-axis is the value of $R_{b,d}$).

Table 8. The summaries of Appendix.

| TEST | 1 | 2 | 3 | 4 | 5 |
|---|---|---|---|---|---|
| $N_{b,d}$ | 25 | 45 | 45 | 45 | 45 |
| $S^*$ | 13824 | $10^5$ | $10^8$ | $10^{12}$ | $10^{15}$ |
| $T_{avg}$ | 0 | 0 | 0.0006 | 0.027 | 3.727119889 |
| $S_{avg}$ | 599.12 | 1203.84 | 28452.16 | 681961.68 | 72578030.4 |
| $S_{avg}/S^*$ | 4.3339120% | 1.2038400% | 0.0284522% | 0.0000682% | 0.0000073% |
| $s_{avg}$ | 26.12 | 40.8 | 225.12 | 739.96 | 3123.28 |
| $R_{avg}$ | 2.0869E-03 | 9.4011E-03 | 9.2658E-05 | 9.1178E-07 | 9.1917E-07 |

1. From Table 8 summarized from Appendix, the larger size of the rework network, the more runtime for the proposed algorithm because rework network reliability is NP-hard and the computational difficulty increases exponentially with the network size [4, 5]. Note that, in Table 8, $N_{b,d}$ is the total number of combinations of different $b$ and $d$ under the related **D**, $S^*$ the total number of possible solutions for each setting of $b$ and $d$ if the proposed algorithm is not implemented to solve the problem, and $\bullet_{avg}$ is the average values of the related $\bullet$.

2. The average runtime $T_{avg}$ is less than 4 computer seconds as shown in Table 8 for each setting of $b$ and $d$ for each test in the worst case. Hence, the proposed BAT is very efficient and it is applicable to solve larger rework problems.

3. The ratio of $S_{avg}/S^*$ is less than 4.5%. Hence, the proposed BAT can reduce the number of solutions to at least 4.5% of the original number. Moreover, the larger size of the rework network, the lower value of $S_{avg}/S^*$. Hence, the above observation further confirms the efficiency of the proposed algorithm.

4. From Appendix, we have $S_{b,d} > S_{b,d+1}$, $S_{b+1,d} > S_{b,d}$, $s_{b,d} > s_{b,d+1}$, and $s_{b+1,d} > s_{b,d}$ because the lower the $d$ and/or the higher the $b$, the greater number of solutions and feasible candidates to send $b$ WIPs into the rework network and receive at least $d$ WIPS without defections in the output for all rework networks.

5. Each diagram in Fig. 8(a), (c), (e), (g), (i) shows the evolution of the phenomenon following the continuous increase in $b$ value. From the original blue curve after reaching its maxima at $b = 2$, i.e., $R_{2,1} < R_{2,2}$ and $R_{2,d} < R_{2,2}$ for all $d > 2$, we observe that not only the following $d$ curves' maxima are on the original decreasing curve, but also those $d$ curves follow the same decreasing path as the original one. Interestingly, the origin of each $d$ curves shares the same $b$ value as its precedent curve maxima.

6. From Fig. 8(b), (d), (f), (h), (j), we have $R_{b,d} > R_{b,d+1}$ and the reason is exactly the same as listed in the 4[th] point. However, we can observe a slow decrease, followed by a dramatic decrease in each $b$ curve, and ultimately reaches its minima as $b = d$ as shown in From Fig. 8(d), (f), (h), (j) due to that $\mathbf{D}(a) = 0.99$ for all $a \in E$ in Eq. (15).

7. Also, from Fig. 8(b), (d), (f), (h), (j), The path followed by each curve substantially depends on the designed input value *b*. As the *d* value varies and the *b* value remains constant, we observe that the variation in each curve is negligible since they are on the same horizontal line, while on the other hand, when the *b* value varies, there will be a noticeable vertical shift in the curve compared to the former.

8. Another important observation is that the more the rework processes, e.g., two in TEST 5, the more runtime we have. For example, the networks are the same in both TEST 4 and TEST 5 if all arcs are ignored. However, the runtime and the number of solutions both are increased exponentially in TEST 5 when compared to TEST 4. The reason is that the more rework processes there are, the more arcs are needed, which also lead to an increase in runtime and in the number of solutions due to the NP-Hard characteristic.

## 7. CONCLUSIONS

A rework network is very commonly found in the industry and manufacturing. However, many studies of the rework network reliability problem are either incorrect in their methods or limited to one rework, being impractical in their assumptions and limitations [23, 27, 28]. To analyze the performance and enhance the applications of a rework network, a more practical general rework network with rework processes is proposed to consider practical phenomena including the one-batch property, preempt property, and deterioration effect.

A new algorithm based on two new multi-state BATs—the top-down deterioration-effect multi-state BAT and the nested multi-state BAT—is proposed to solve the proposed new rework network reliability problem efficiently and systematically. To increase the efficiency, the proposed algorithm splits the proposed rework problem into disjoint production lines, including perfect and rework production lines. Each individual production line has its own top-down deterioration-effect multi-state BAT to generate related feasible deterioration-effect multi-state vectors and solve the proposed rework problem systematically. Subsequently, the proposed nested multi-state BAT is implemented together with the node and split constraints to combine these deterioration-effect multi-state vectors into feasible solutions. The probabilities of all the feasible solutions are calculated and added to yield the reliability of the rework problem.

The experiments validate the performance of the proposed algorithm for solving the one-batch preempt multi-state multi-rework network reliability problem. It enables the extension of the proposed algorithm to solve different types of rework problems, e.g., including fuzzy to consider the uncertainty in real-life applications and the multi-objective function without limiting the reliability only, with increasing the scale size.

**ACKNOWLEDGMENT**

This research was supported in part by the Ministry of Science and Technology of Taiwan (MOST 107-2221-E-007-072-MY3). This article was once submitted to arXiv as a temporary submission that was just for reference and did not provide the copyright.

**APPENDIX**

Table A1. Results of TEST 1 ($n=2$ and $m=6$).

| $b$ | $d$ | $m_1$ | $m_2$ | T | $S_{b,d}$ | $s_{b,d}$ | $R_{b,d}$ |
|---|---|---|---|---|---|---|---|
| 1 | 1 | 2 | 4 | 0 | 8 | 1 | 1.0692E-05 |
| 2 | 1 | 5 | 10 | 0 | 50 | 5 | 2.95585E-05 |
| 2 | 2 | 3 | 10 | 0 | 30 | 1 | 2.54042E-05 |
| 3 | 1 | 9 | 20 | 0 | 180 | 15 | 6.19411E-05 |
| 3 | 2 | 7 | 20 | 0 | 140 | 5 | 6.05212E-05 |
| 3 | 3 | 4 | 20 | 0 | 80 | 1 | 4.34595E-05 |
| 4 | 1 | 14 | 35 | 0 | 490 | 36 | 0.000629956 |
| 4 | 2 | 12 | 35 | 0 | 420 | 16 | 0.000629356 |
| 4 | 3 | 9 | 35 | 0 | 315 | 5 | 0.000618127 |
| 4 | 4 | 5 | 35 | 0 | 175 | 1 | 0.000542113 |
| 5 | 1 | 20 | 56 | 0 | 1120 | 74 | 0.00928899 |
| 5 | 2 | 18 | 56 | 0 | 1008 | 39 | 0.009288843 |
| 5 | 3 | 15 | 56 | 0 | 840 | 16 | 0.009285097 |
| 5 | 4 | 11 | 56 | 0 | 616 | 5 | 0.009235384 |
| 5 | 5 | 6 | 56 | 0 | 336 | 1 | 0.008280393 |
| 6 | 1 | 20 | 62 | 0 | 1240 | 108 | 0.00087589 |
| 6 | 2 | 18 | 62 | 0 | 1116 | 59 | 0.000875874 |
| 6 | 3 | 15 | 62 | 0 | 930 | 24 | 0.000875377 |
| 6 | 4 | 11 | 62 | 0 | 682 | 7 | 0.000858893 |
| 6 | 5 | 6 | 62 | 0 | 372 | 1 | 0.000496823 |
| 7 | 1 | 20 | 69 | 0 | 1380 | 128 | 3.78339E-05 |
| 7 | 2 | 18 | 69 | 0 | 1242 | 70 | 3.78313E-05 |
| 7 | 3 | 15 | 69 | 0 | 1035 | 27 | 3.76247E-05 |
| 7 | 4 | 11 | 69 | 0 | 759 | 7 | 2.99841E-05 |
| 7 | 5 | 6 | 69 | 0 | 414 | 1 | 1.73888E-05 |

**Table A2.** Results of TEST 2 ($n=2$ and $m=6$).

| $b$ | $d$ | $m_1$ | $m_2$ | T | $S_{b,d}$ | $s_{b,d}$ | $R_{b,d}$ |
|---|---|---|---|---|---|---|---|
| 1 | 1 | 2 | 4 | 0 | 8 | 1 | 9.70299E-03 |
| 2 | 1 | 5 | 10 | 0 | 50 | 5 | 9.80482E-03 |
| 2 | 2 | 3 | 10 | 0 | 30 | 1 | 9.41480E-03 |
| 3 | 1 | 9 | 20 | 0 | 180 | 15 | 9.71735E-03 |
| 3 | 2 | 7 | 20 | 0 | 140 | 5 | 9.70563E-03 |
| 3 | 3 | 4 | 20 | 0 | 80 | 1 | 9.13517E-03 |
| 4 | 1 | 14 | 35 | 0 | 490 | 36 | 9.62867E-03 |
| 4 | 2 | 12 | 35 | 0 | 420 | 16 | 9.62835E-03 |
| 4 | 3 | 9 | 35 | 0 | 315 | 5 | 9.60548E-03 |
| 4 | 4 | 5 | 35 | 0 | 175 | 1 | 8.86385E-03 |
| 5 | 1 | 20 | 56 | 0 | 1120 | 74 | 9.54242E-03 |
| 5 | 2 | 18 | 56 | 0 | 1008 | 39 | 9.54241E-03 |
| 5 | 3 | 15 | 56 | 0 | 840 | 16 | 9.54165E-03 |
| 5 | 4 | 11 | 56 | 0 | 616 | 5 | 9.50448E-03 |
| 5 | 5 | 6 | 56 | 0 | 336 | 1 | 8.60059E-03 |
| 6 | 1 | 27 | 84 | 0 | 2268 | 138 | 9.45858E-03 |
| 6 | 2 | 25 | 84 | 0 | 2100 | 82 | 9.45858E-03 |
| 6 | 3 | 22 | 84 | 0 | 1848 | 40 | 9.45856E-03 |
| 6 | 4 | 18 | 84 | 0 | 1512 | 16 | 9.45707E-03 |
| 6 | 5 | 13 | 84 | 0 | 1092 | 5 | 9.40270E-03 |
| 6 | 6 | 7 | 84 | 0 | 588 | 1 | 8.34514E-03 |
| 7 | 1 | 35 | 120 | 0 | 4200 | 238 | 9.37705E-03 |
| 7 | 2 | 33 | 120 | 0 | 3960 | 154 | 9.37705E-03 |
| 7 | 3 | 30 | 120 | 0 | 3600 | 85 | 9.37705E-03 |
| 7 | 4 | 26 | 120 | 0 | 3120 | 40 | 9.37700E-03 |
| 7 | 5 | 21 | 120 | 0 | 2520 | 16 | 9.37447E-03 |
| 7 | 6 | 15 | 120 | 0 | 1800 | 5 | 9.30024E-03 |
| 7 | 7 | 8 | 120 | 0 | 960 | 1 | 8.09728E-03 |
| 8 | 1 | 44 | 165 | 0 | 7260 | 388 | 9.29773E-03 |
| 8 | 2 | 42 | 165 | 0 | 6930 | 268 | 9.29773E-03 |
| 8 | 3 | 39 | 165 | 0 | 6435 | 162 | 9.29773E-03 |
| 8 | 4 | 35 | 165 | 0 | 5775 | 86 | 9.29773E-03 |
| 8 | 5 | 30 | 165 | 0 | 4950 | 40 | 9.29763E-03 |
| 8 | 6 | 24 | 165 | 0 | 3960 | 16 | 9.29368E-03 |
| 8 | 7 | 17 | 165 | 0 | 2805 | 5 | 9.19717E-03 |
| 8 | 8 | 9 | 165 | 0 | 1485 | 1 | 7.85678E-03 |
| 9 | 1 | 54 | 220 | 0 | 11880 | 603 | 9.22052E-03 |
| 9 | 2 | 52 | 220 | 0 | 11440 | 438 | 9.22052E-03 |
| 9 | 3 | 49 | 220 | 0 | 10780 | 284 | 9.22052E-03 |
| 9 | 4 | 45 | 220 | 0 | 9900 | 165 | 9.22052E-03 |
| 9 | 5 | 40 | 220 | 0 | 8800 | 86 | 9.22052E-03 |
| 9 | 6 | 34 | 220 | 0 | 7480 | 40 | 9.22034E-03 |
| 9 | 7 | 27 | 220 | 0 | 5940 | 16 | 9.21458E-03 |
| 9 | 8 | 19 | 220 | 0 | 4180 | 5 | 9.09358E-03 |
| 9 | 9 | 10 | 220 | 0 | 2200 | 1 | 7.62343E-03 |

**Table A3.** Results of TEST 3 ($n=4$ and $m=9$).

| $b$ | $d$ | $m_1$ | $m_2$ | T | $S_{b,d}$ | $s_{b,d}$ | $R_{b,d}$ |
|---|---|---|---|---|---|---|---|
| 1 | 1 | 4 | 5 | 0 | 20 | 1 | 9.50990E-05 |
| 2 | 1 | 14 | 15 | 0 | 210 | 8 | 9.79889E-05 |
| 2 | 2 | 10 | 15 | 0 | 150 | 1 | 9.04382E-05 |
| 3 | 1 | 34 | 35 | 0 | 1190 | 36 | 9.72801E-05 |
| 3 | 2 | 30 | 35 | 0 | 1050 | 8 | 9.68302E-05 |
| 3 | 3 | 20 | 35 | 0 | 700 | 1 | 8.60059E-05 |
| 4 | 1 | 69 | 70 | 0 | 4830 | 123 | 9.64683E-05 |
| 4 | 2 | 65 | 70 | 0 | 4550 | 39 | 9.64446E-05 |
| 4 | 3 | 55 | 70 | 0 | 3850 | 8 | 9.55836E-05 |
| 4 | 4 | 35 | 70 | 0 | 2450 | 1 | 8.17907E-05 |
| 5 | 1 | 125 | 126 | 0 | 15750 | 346 | 9.56899E-05 |
| 5 | 2 | 121 | 126 | 0 | 15246 | 136 | 9.56887E-05 |
| 5 | 3 | 111 | 126 | 0 | 13986 | 39 | 9.56319E-05 |
| 5 | 4 | 91 | 126 | 0 | 11466 | 8 | 9.42589E-05 |
| 5 | 5 | 56 | 126 | 0 | 7056 | 1 | 7.77822E-05 |
| 6 | 1 | 209 | 210 | 0 | 43890 | 855 | 9.49479E-05 |
| 6 | 2 | 205 | 210 | 0 | 43050 | 393 | 9.49479E-05 |
| 6 | 3 | 195 | 210 | 0 | 40950 | 140 | 9.49445E-05 |
| 6 | 4 | 175 | 210 | 0 | 36750 | 39 | 9.48357E-05 |
| 6 | 5 | 140 | 210 | 0 | 29400 | 8 | 9.28651E-05 |
| 6 | 6 | 84 | 210 | 0 | 17640 | 1 | 7.39701E-05 |
| 7 | 1 | 329 | 330 | 0.015 | 108570 | 1905 | 9.42403E-05 |
| 7 | 2 | 325 | 330 | 0 | 107250 | 981 | 9.42403E-05 |
| 7 | 3 | 315 | 330 | 0 | 103950 | 410 | 9.42401E-05 |
| 7 | 4 | 295 | 330 | 0 | 97350 | 140 | 9.42326E-05 |
| 7 | 5 | 260 | 330 | 0.016 | 85800 | 39 | 9.40503E-05 |
| 7 | 6 | 204 | 330 | 0 | 67320 | 8 | 9.14106E-05 |
| 7 | 7 | 120 | 330 | 0 | 39600 | 1 | 7.03448E-05 |
| 8 | 1 | 494 | 495 | 0 | 244530 | 3926 | 9.35649E-05 |
| 8 | 2 | 490 | 495 | 0.015 | 242550 | 2210 | 9.35649E-05 |
| 8 | 3 | 480 | 495 | 0 | 237600 | 1041 | 9.35649E-05 |
| 8 | 4 | 460 | 495 | 0.016 | 227700 | 415 | 9.35644E-05 |
| 8 | 5 | 425 | 495 | 0 | 210375 | 140 | 9.35500E-05 |
| 8 | 6 | 369 | 495 | 0.016 | 182655 | 39 | 9.32710E-05 |
| 8 | 7 | 285 | 495 | 0 | 141075 | 8 | 8.99033E-05 |
| 8 | 8 | 165 | 495 | 0 | 81675 | 1 | 6.68972E-05 |
| 9 | 1 | 714 | 715 | 0.015 | 510510 | 7578 | 9.29198E-05 |
| 9 | 2 | 710 | 715 | 0.016 | 507650 | 4575 | 9.29198E-05 |
| 9 | 3 | 700 | 715 | 0.018 | 500500 | 2368 | 9.29198E-05 |
| 9 | 4 | 680 | 715 | 0.016 | 486200 | 1062 | 9.29198E-05 |
| 9 | 5 | 645 | 715 | 0.015 | 461175 | 415 | 9.29188E-05 |
| 9 | 6 | 589 | 715 | 0 | 421135 | 140 | 9.28940E-05 |
| 9 | 7 | 505 | 715 | 0.016 | 361075 | 39 | 9.24935E-05 |
| 9 | 8 | 385 | 715 | 0 | 275275 | 8 | 8.83506E-05 |
| 9 | 9 | 220 | 715 | 0.018 | 157300 | 1 | 6.36186E-05 |

**Table A4.** Results of TEST 4 ($n=6$ and $m=13$).

| $b$ | $d$ | $m_1$ | $m_2$ | T | $S_{b,d}$ | $s_{b,d}$ | $R_{b,d}$ |
|---|---|---|---|---|---|---|---|
| 1 | 1 | 6 | 7 | 0 | 42 | 1 | 9.32065E-07 |
| 2 | 1 | 27 | 28 | 0 | 756 | 10 | 9.78019E-07 |
| 2 | 2 | 21 | 28 | 0 | 588 | 1 | 8.68746E-07 |
| 3 | 1 | 83 | 84 | 0 | 6972 | 59 | 9.72155E-07 |
| 3 | 2 | 77 | 84 | 0 | 6468 | 10 | 9.62508E-07 |
| 3 | 3 | 56 | 84 | 0 | 4704 | 1 | 8.09728E-07 |
| 4 | 1 | 209 | 210 | 0.002 | 43890 | 259 | 9.63336E-07 |
| 4 | 2 | 203 | 210 | 0.002 | 42630 | 62 | 9.62579E-07 |
| 4 | 3 | 182 | 210 | 0.001 | 38220 | 10 | 9.44593E-07 |
| 4 | 4 | 126 | 210 | 0 | 26460 | 1 | 7.54720E-07 |
| 5 | 1 | 461 | 462 | 0.01 | 212982 | 930 | 9.54436E-07 |
| 5 | 2 | 455 | 462 | 0.01 | 210210 | 278 | 9.54380E-07 |
| 5 | 3 | 434 | 462 | 0.008 | 200508 | 62 | 9.52614E-07 |
| 5 | 4 | 378 | 462 | 0.008 | 174636 | 10 | 9.24674E-07 |
| 5 | 5 | 252 | 462 | 0.005 | 116424 | 1 | 7.03448E-07 |
| 6 | 1 | 923 | 924 | 0.037 | 852852 | 2888 | 9.45621E-07 |
| 6 | 2 | 917 | 924 | 0.032 | 847308 | 1020 | 9.45617E-07 |
| 6 | 3 | 896 | 924 | 0.03 | 827904 | 282 | 9.45461E-07 |
| 6 | 4 | 840 | 924 | 0.028 | 776160 | 62 | 9.42170E-07 |
| 6 | 5 | 714 | 924 | 0.023 | 659736 | 10 | 9.03104E-07 |
| 6 | 6 | 462 | 924 | 0.015 | 426888 | 1 | 6.55660E-07 |
| 7 | 1 | 1715 | 1716 | 0.132 | 2942940 | 8005 | 9.36901E-07 |
| 7 | 2 | 1709 | 1716 | 0.116 | 2932644 | 3209 | 9.36901E-07 |
| 7 | 3 | 1688 | 1716 | 0.112 | 2896608 | 1045 | 9.36888E-07 |
| 7 | 4 | 1632 | 1716 | 0.104 | 2800512 | 282 | 9.36549E-07 |
| 7 | 5 | 1506 | 1716 | 0.092 | 2584296 | 62 | 9.31179E-07 |
| 7 | 6 | 1254 | 1716 | 0.077 | 2151864 | 10 | 8.80201E-07 |
| 7 | 7 | 792 | 1716 | 0.048 | 1359072 | 1 | 6.11118E-07 |
| 8 | 1 | 3002 | 3003 | 0.368 | 9015006 | 20273 | 9.28273E-07 |
| 8 | 2 | 2996 | 3003 | 0.358 | 8996988 | 8997 | 9.28273E-07 |
| 8 | 3 | 2975 | 3003 | 0.365 | 8933925 | 3325 | 9.28272E-07 |
| 8 | 4 | 2919 | 3003 | 0.346 | 8765757 | 1050 | 9.28240E-07 |
| 8 | 5 | 2793 | 3003 | 0.324 | 8387379 | 282 | 9.27608E-07 |
| 8 | 6 | 2541 | 3003 | 0.265 | 7630623 | 62 | 9.19600E-07 |
| 8 | 7 | 2079 | 3003 | 0.224 | 6243237 | 10 | 8.56243E-07 |
| 8 | 8 | 1287 | 3003 | 0.138 | 3864861 | 1 | 5.69602E-07 |
| 9 | 1 | 5004 | 5005 | 1.003 | 25045020 | 47638 | 9.19737E-07 |
| 9 | 2 | 4998 | 5005 | 1.022 | 25014990 | 22964 | 9.19737E-07 |
| 9 | 3 | 4977 | 5005 | 0.989 | 24909885 | 9404 | 9.19737E-07 |
| 9 | 4 | 4921 | 5005 | 0.962 | 24629605 | 3356 | 9.19734E-07 |
| 9 | 5 | 4795 | 5005 | 0.858 | 23998975 | 1050 | 9.19667E-07 |
| 9 | 6 | 4543 | 5005 | 0.823 | 22737715 | 282 | 9.18606E-07 |
| 9 | 7 | 4081 | 5005 | 0.738 | 20425405 | 62 | 9.07409E-07 |
| 9 | 8 | 3289 | 5005 | 0.601 | 16461445 | 10 | 8.31481E-07 |
| 9 | 9 | 2002 | 5005 | 0.373 | 10020010 | 1 | 5.30906E-07 |

**Table A5.** Results of TEST 5 ($n=6$ and $m=16$).

| $b$ | $d$ | $m_1$ | $m_2$ | $m_3$ | T | $S_{b,d}$ | $s_{b,d}$ | $R_{b,d}$ |
|---|---|---|---|---|---|---|---|---|
| 1 | 1 | 6 | 7 | 4 | 0 | 168 | 1 | 9.32065E-07 |
| 2 | 1 | 27 | 28 | 10 | 0 | 7560 | 17 | 9.79463E-07 |
| 2 | 2 | 21 | 28 | 10 | 0 | 5880 | 1 | 8.68746E-07 |
| 3 | 1 | 83 | 84 | 20 | 0.015 | 139440 | 127 | 9.76136E-07 |
| 3 | 2 | 77 | 84 | 20 | 0 | 129360 | 19 | 9.66306E-07 |
| 3 | 3 | 56 | 84 | 20 | 0 | 94080 | 1 | 8.09728E-07 |
| 4 | 1 | 209 | 210 | 35 | 0.092 | 1536150 | 687 | 9.70879E-07 |
| 4 | 2 | 203 | 210 | 35 | 0.083 | 1492050 | 172 | 9.70105E-07 |
| 4 | 3 | 182 | 210 | 35 | 0.097 | 1337700 | 19 | 9.51539E-07 |
| 4 | 4 | 126 | 210 | 35 | 0.053 | 926100 | 1 | 7.54720E-07 |
| 5 | 1 | 461 | 462 | 56 | 0.681 | 11926992 | 2930 | 9.66481E-07 |
| 5 | 2 | 455 | 462 | 56 | 0.627 | 11771760 | 967 | 9.66424E-07 |
| 5 | 3 | 434 | 462 | 56 | 0.597 | 11228448 | 186 | 9.64596E-07 |
| 5 | 4 | 378 | 462 | 56 | 0.597 | 9779616 | 19 | 9.35373E-07 |
| 5 | 5 | 252 | 462 | 56 | 0.361 | 6519744 | 1 | 7.03448E-07 |
| 6 | 1 | 923 | 924 | 84 | 3.986 | 71639568 | 10681 | 9.63018E-07 |
| 6 | 2 | 917 | 924 | 84 | 4.138 | 71173872 | 4286 | 9.63014E-07 |
| 6 | 3 | 896 | 924 | 84 | 3.659 | 69543936 | 1158 | 9.62852E-07 |
| 6 | 4 | 840 | 924 | 84 | 3.464 | 65197440 | 189 | 9.59399E-07 |
| 6 | 5 | 714 | 924 | 84 | 3.344 | 55417824 | 19 | 9.18003E-07 |
| 6 | 6 | 462 | 924 | 84 | 1.868 | 35858592 | 1 | 6.55660E-07 |
| 7 | 1 | 1715 | 1716 | 120 | 18.738 | 353152800 | 34258 | 9.60415E-07 |
| 7 | 2 | 1709 | 1716 | 120 | 18.793 | 351917280 | 15789 | 9.60415E-07 |
| 7 | 3 | 1688 | 1716 | 120 | 19.056 | 347592960 | 5324 | 9.60401E-07 |
| 7 | 4 | 1632 | 1716 | 120 | 18.465 | 336061440 | 1229 | 9.60045E-07 |
| 7 | 5 | 1506 | 1716 | 120 | 20.63 | 310115520 | 189 | 9.54337E-07 |
| 7 | 6 | 1254 | 1716 | 120 | 16.971 | 258223680 | 19 | 8.99607E-07 |
| 7 | 7 | 792 | 1716 | 120 | 9.54 | 163088640 | 1 | 6.11118E-07 |
| 8 | 1 | 3002 | 3003 | 165 | 80.767 | 1487475990 | 99576 | 9.58596E-07 |
| 8 | 2 | 2996 | 3003 | 165 | 78.61 | 1484503020 | 51109 | 9.58596E-07 |
| 8 | 3 | 2975 | 3003 | 165 | 85.857 | 1474097625 | 20241 | 9.58595E-07 |
| 8 | 4 | 2919 | 3003 | 165 | 79.563 | 1446349905 | 6003 | 9.58561E-07 |
| 8 | 5 | 2793 | 3003 | 165 | 71.1 | 1383917535 | 1249 | 9.57888E-07 |
| 8 | 6 | 2541 | 3003 | 165 | 76.965 | 1259052795 | 189 | 9.49262E-07 |
| 8 | 7 | 2079 | 3003 | 165 | 51.568 | 1030134105 | 19 | 8.80350E-07 |
| 8 | 8 | 1287 | 3003 | 165 | 31.523 | 637702065 | 1 | 5.69602E-07 |
| 9 | 1 | 5004 | 5005 | 220 | 303.246 | 5509904400 | 266268 | 9.57489E-07 |
| 9 | 2 | 4998 | 5005 | 220 | 311.84 | 5503297800 | 148680 | 9.57489E-07 |
| 9 | 3 | 4977 | 5005 | 220 | 303.592 | 5480174700 | 66473 | 9.57489E-07 |
| 9 | 4 | 4921 | 5005 | 220 | 304.713 | 5418513100 | 23526 | 9.57486E-07 |
| 9 | 5 | 4795 | 5005 | 220 | 297.514 | 5279774500 | 6283 | 9.57415E-07 |
| 9 | 6 | 4543 | 5005 | 220 | 276.028 | 5002297300 | 1253 | 9.56271E-07 |
| 9 | 7 | 4081 | 5005 | 220 | 241.966 | 4493589100 | 189 | 9.44048E-07 |
| 9 | 8 | 3289 | 5005 | 220 | 192.281 | 3621517900 | 19 | 8.60379E-07 |
| 9 | 9 | 2002 | 5005 | 220 | 121.658 | 2204402200 | 1 | 5.30906E-07 |

Table A6[*]. Comparisons between the rework network in Fig. 4 and that in Fig. 5.

| b | d | $T_5-T_4$ | $T_5/T_4$ | $S_{b,d,5}/S_{b,d,4}$ | $s_{b,d,5}/s_{b,d,4}$ | $R_{b,d,5}/R_{b,d,4}$ |
|---|---|---|---|---|---|---|
| 1 | 1 | 0 | | 4.0 | 1 | 1.00000000000 |
| 2 | 1 | 0 | | 10.0 | 1.7 | 1.00110748410 |
| 2 | 2 | 0 | | 10.0 | 1 | 1.00000000000 |
| 3 | 1 | 0.015 | | 20.0 | 2.152542 | 1.00300764606 |
| 3 | 2 | 0 | | 20.0 | 1.9 | 1.00289772893 |
| 3 | 3 | 0 | | 20.0 | 1 | 1.00000000000 |
| 4 | 1 | 0.092 | | 35.0 | 2.65251 | 1.00568292988 |
| 4 | 2 | 0.083 | | 35.0 | 2.774194 | 1.00567485376 |
| 4 | 3 | 0.097 | | 35.0 | 1.9 | 1.00535612843 |
| 4 | 4 | 0.053 | | 35.0 | 1 | 1.00000000000 |
| 5 | 1 | 0.666 | 45.4 | 56.0 | 3.150538 | 1.00908208696 |
| 5 | 2 | 0.627 | | 56.0 | 3.478417 | 1.00908160389 |
| 5 | 3 | 0.581 | 37.3 | 56.0 | 3 | 1.00905422068 |
| 5 | 4 | 0.597 | | 56.0 | 1.9 | 1.00839226109 |
| 5 | 5 | 0.361 | | 56.0 | 1 | 1.00000000000 |
| 6 | 1 | 3.955 | 128.6 | 84.0 | 3.698407 | 1.01314628299 |
| 6 | 2 | 4.091 | 88.0 | 84.0 | 4.201961 | 1.01314626085 |
| 6 | 3 | 3.628 | 118.0 | 84.0 | 4.106383 | 1.01314435655 |
| 6 | 4 | 3.433 | 111.7 | 84.0 | 3.048387 | 1.01307895283 |
| 6 | 5 | 3.325 | 176.0 | 84.0 | 1.9 | 1.01193147947 |
| 6 | 6 | 1.853 | 124.5 | 84.0 | 1 | 1.00000000000 |
| 7 | 1 | 18.613 | 149.9 | 120.0 | 4.279575 | 1.01782001175 |
| 7 | 2 | 18.668 | 150.3 | 120.0 | 4.920224 | 1.01782001131 |
| 7 | 3 | 18.946 | 173.2 | 120.0 | 5.094737 | 1.01781990907 |
| 7 | 4 | 18.355 | 167.9 | 120.0 | 4.358156 | 1.01781473200 |
| 7 | 5 | 20.536 | 219.5 | 120.0 | 3.048387 | 1.01768702021 |
| 7 | 6 | 16.893 | 217.6 | 120.0 | 1.9 | 1.01591161160 |
| 7 | 7 | 9.493 | 203.0 | 120.0 | 1 | 1.00000000000 |
| 8 | 1 | 80.401 | 220.7 | 165.0 | 4.911755 | 1.02305169045 |
| 8 | 2 | 78.266 | 228.5 | 165.0 | 5.680671 | 1.02305169050 |
| 8 | 3 | 85.527 | 260.2 | 165.0 | 6.087519 | 1.02305168762 |
| 8 | 4 | 79.232 | 240.4 | 165.0 | 5.717143 | 1.02305137445 |
| 8 | 5 | 70.804 | 240.2 | 165.0 | 4.429078 | 1.02304008533 |
| 8 | 6 | 76.697 | 287.2 | 165.0 | 3.048387 | 1.02282148692 |
| 8 | 7 | 51.347 | 233.3 | 165.0 | 1.9 | 1.02028045694 |
| 8 | 8 | 31.382 | 223.6 | 165.0 | 1 | 1.00000000000 |
| 9 | 1 | 302.289 | 316.9 | 220.0 | 5.589403 | 1.02879348546 |
| 9 | 2 | 310.866 | 320.2 | 220.0 | 6.474482 | 1.02879348547 |
| 9 | 3 | 302.574 | 298.2 | 220.0 | 7.068588 | 1.02879348569 |
| 9 | 4 | 303.730 | 310.0 | 220.0 | 7.010131 | 1.02879347558 |
| 9 | 5 | 296.363 | 258.5 | 220.0 | 5.98381 | 1.02879272538 |
| 9 | 6 | 275.018 | 273.3 | 220.0 | 4.443262 | 1.02877149087 |
| 9 | 7 | 241.122 | 286.7 | 220.0 | 3.048387 | 1.02843057376 |
| 9 | 8 | 191.622 | 291.8 | 220.0 | 1.9 | 1.02499386225 |
| 9 | 9 | 121.283 | 324.4 | 220.0 | 1 | 1.00000000000 |

[*]$\bullet_i$: the value of $\bullet$ for Fig. $i$, e.g., $T_5$ is the related runtime of the rework network in Fig. 5.